\documentclass[twocolumn,showpacs,preprintnumbers,amsmath,amssymb]{revtex4}

\usepackage{graphicx}
\usepackage{dcolumn}
\usepackage{bm}% bold math

\newcommand{\AsGa}{{\ensuremath{\mathrm{As_{Ga}}\,}}}
\newcommand{\Asi}{{\ensuremath{\mathrm{As_{i}}\,}}}
\newcommand{\VGa}{{\ensuremath{\mathrm{V_{Ga}}\,}}}
\newcommand{\ELt}{{\ensuremath{\mathrm{EL2^{0}}\,}}}
\newcommand{\ELts}{{\ensuremath{\mathrm{EL2^{*}}\,}}}
\newcommand{\Ctv}{{\ensuremath{\mathrm{C_{3v}}\,}}}
\newcommand{\Td}{{\ensuremath{\mathrm{T_d}\,}}}
\newcommand{\lDOS}{$\ell$DOS}

\begin{document}

\title{The electronic structure around As antisite near (110) surface of GaAs}

\author{Yusuke Iguchi}
\author{Takeo Fujiwara}
\author{Akira Hida}
\author{Koji Maeda}
\affiliation{Department of Applied Physics, The University of Tokyo, Tokyo 113-8656, Japan}

\date{\today}
%======================================================================%
%                             Abstract                                 %
%======================================================================%
\begin{abstract}
 The electronic structure around a single As antisite in GaAs 
is investigated in bulk and near the surface
both in the stable and the metastable atomic configurations. 
 The most characteristic electronic structures of As antisite 
 is the existence of the localized p-orbitals
 extending from the As antisite.
  The major component of the highest occupied state on As antisite
in the stable configuration  
is s-orbital connecting with neighboring As atoms with nodes 
whereas that in the metastable configuration 
is p-orbital connecting without nodes. 
 Localized p-orbitals on the surrounding As atoms around the As antisite
exist in every configuration of As antisite.
Such features are retained  except 
the case of the As antisite located just in the surface layer
in which the midgap level is smeared into the conduction band 
and  no localized states exist near the top of the valence band.
 Scanning tunneling microscopic images of defects observed 
in low-temperature grown GaAs, possibly assigned as As antisite, 
the origin of the metastability,
and the peculiarity of the defects 
in the surface layer are discussed.
\end{abstract}

%\keywords{EL2 STM}
\maketitle

%%%%%%%%%%%%%%%%%%%%%%%%%%%%%%%%%%%%%%%%%%%%%%%%%%%%%%%%%%%%%%%%%%%%%%%
%                        Introduction                                 %
%%%%%%%%%%%%%%%%%%%%%%%%%%%%%%%%%%%%%%%%%%%%%%%%%%%%%%%%%%%%%%%%%%%%%%%
\section{\label{sec:introduction}Introduction}
Bulk GaAs crystals grown under As-rich conditions are known
to contain As-related defects called EL2 centers \cite{kaminska1993}
that form a deep double donor level at the middle of the band gap.
The EL2 centers compensate residual shallow acceptors 
thus giving the crystals a semi-insulating property useful 
for device applications. 
The EL2 centers induce a characteristic optical absorption
extending to the sub-band gap region with a spectral hump
peaking around 1.2 eV.
The most peculiar feature of the EL2 centers is
the photoquenching effect 
in which the EL2-originated optical absorption becomes diminished 
when the crystal is optically illuminated 
with an intra-center excitation
at low temperatures.~\cite{martin1981}
The centers once quenched are, however, recovered 
if the crystal is heated or illuminated 
with a different infrared light.~\cite{fischer1987}
This reversibility and the lost of the effect at high temperatures 
indicate that the EL2 centers, when they are electrically neutral, 
are transformable between the stable (\ELt) 
and the metastable (\ELts) state.

For the atomic structure of EL2 centers,
the most commonly accepted model is isolated arsenic antisite defects
 (\AsGa).
First principle calculations \cite{chadi1988, dabrowski1988}
based on the \AsGa model showed
that the metastable \ELts may be a close pair of
an interstitial arsenic (\Asi) and a gallium vacancy (\VGa) 
that is formed by the displacement of the antisite As atom
from the lattice point 
to a puckered interstitial position in a [111] direction 
leaving a \VGa\ behind. 
The point symmetry lowering from \Td to \Ctv
by the displacement 
was supported by optical absorption experiment.~\cite{trautman1992}
However, the atomic structure of \ELt and \ELts
have been controversial since some defects,
though exhibiting features of EL2 centers, 
were found to show no photoquenching effect.~\cite{omling1986, manasreh1990}
The presence of such variations in EL2-like centers arose 
an argument that the EL2 centers form a `family' 
including various structures rather 
than have a unique atomic configuration.~\cite{taniguchi1983}

Low-temperature grown (LT-) GaAs crystals 
homoepitaxially grown under an excess As pressure 
contain a high density of As-related defects. 
Feenstra and his coworkers~\cite{feenstra1993} applied 
scanning tunneling microscopy (STM) 
to atomic level observations of 
individual point defects abundantly found in LT-GaAs epifilms. 
Four different defect contrasts observed in different sizes 
were attributed to \AsGa atoms located in different depths 
from the sample surface. 
Recently, some of the present authors \cite{hida2001} provided 
direct evidence for the defect being EL2: 
They found that the STM contrasts of the defects in LT-GaAs 
samples kept at a low temperature (90K) 
are drastically changed by light illumination with an excitation 
spectrum nearly identical to that 
for the photoquenching effect of EL2 centers
in bulk crystal. 
The local density of states (\lDOS) measured 
by scanning tunneling spectroscopy (STS) 
at the defect sites shows 
that the donor gap state present in the normal state 
disappears in the metastable state, 
conforming the characteristics of EL2.

Theoretically simulated STM images of \AsGa defects 
show an apparently good agreement with experimental images 
as far as the defects 
in the second largest contrast~\cite{feenstra1993} 
are concerned.~\cite{capaz1995, zhang1999} 
However, no systematic comparison has been made 
for the STM image contrasts in other sizes 
and those in the metastable state. \cite{footnote1}

In this paper, we will report results 
of first principle electronic structure calculations  
of an \AsGa in GaAs bulk crystal 
and those near the surface, both in the stable and the metastable state. 
The main aim of the present study is to answer a na{\" \i}ve question 
if defects near the surface that can be probed by STM 
may be substantially different or not in the structures 
and physical properties from the defects in the bulk crystals. 
In Sec.~\ref{sec:comp}, we describe the methods of calculations 
based on the atomic structure models. 
In Sec.~\ref{sec:elec}, we show the electronic structures 
of the highest occupied level in real space 
and discuss the nature of the level which differs depending 
on the depth of the \AsGa from the surface. 
Section \ref{sec:discussion} is devoted to a comparison 
between the simulated STM images and experimental ones 
and discussions about the effects of surface 
on the properties of \AsGa centers in light 
of the calculated electronic structures. 
In Sec.~\ref{sec:conclusion}, we conclude this paper.

%%%%%%%%%%%%%%%%%%%%%%%%%%%%%%%%%%%%%%%%%%%%%%%%%%%%%%%%%%%%%%%%%%%%%%%
%                     Computational Details                           %
%%%%%%%%%%%%%%%%%%%%%%%%%%%%%%%%%%%%%%%%%%%%%%%%%%%%%%%%%%%%%%%%%%%%%%%
\section{\label{sec:comp} Computational Details}

Since a considerable lattice expansion is present 
around \AsGa\ defects 
due to the antibonding nature of 
the double donor level in the band gap, \cite{bonapasta2000} 
it is needed to calculate the electronic structure 
allowing lattice relaxation with sufficient accuracy.
 For this purpose, we used the density-functional (DFT) theory 
with the local-density approximation (LDA) employing 
two different methods for electronic structure calculations.
 One is the method using 
the norm-conserving pseudopotential of the Troullier-Martin type 
with the plane wave basis set \cite{troullier1991} 
and the other is the tight-binding Linear Muffin-Tin Orbital (LMTO) method. 
\cite{andersen1984}

In the case of the norm-conserving pseudopotential of 
the Troullier-Martin type, 
we adopt a supercell and 
the cut-off energy for the plane wave basis was set to be 10~Rydberg.
The ${\bf k}$-point sampling in the Brillouin zone 
was done only at the $\Gamma$ point.
%
%By the norm-conserving pseudopotential method with plane wave bases,
%one can calculate the forces 
%acting on atoms easily so that this method
%is often used for molecular dynamics, 
%phonon or lattice relaxation calculations.
%
We relaxed atomic positions, 
and calculated the electron density with the plane wave basis.

The LMTO basis is localized around each atom 
and hence we can analyze the contribution of the atomic orbital 
to a specific energy state in concern. 
The LMTO basis set is minimal 
and each basis function can be described 
by a small number of parameters 
so that we can perform calculations 
with relatively small memory and computing time.

The procedure for calculating the electronic structures 
in the model system that is bounded with a surface and contains 
an \AsGa defect is as follows:\\
(1) Relaxing a relatively small lattice system with an \AsGa 
and a surface by the pseudopotential method.\\
(2) Embedding the relaxed small system in a larger perfect lattice 
with the atomic configuration around the \AsGa unchanged.\\
(3) Calculating the electron density with $\Gamma$ point 
in a supercell by the pseudopotential method, and other properties 
such as the $E - {\bf k}$ relation, 
the total density of states (DOS) and local density of states 
(\lDOS) of each atom by the tight-binding LMTO method.

For \AsGa\ defects in bulk crystals, 
we introduced an \AsGa\ defect by replacing a Ga atom 
with an As atom in a $2\times 2\times 2$
cubic unit cells containing 64 atoms.
These atoms were then subject to relaxation 
by means of the molecular dynamics with plane wave bases.
The lattice constant was fixed to the experimental value 5.654 \AA\ 
during the relaxation.
We found that the distances between the \AsGa and the neighboring atoms 
after relaxation are similar 
to the previous results. \cite{kaxiras1989, bonapasta2000}

 For calculations of the defect structure in the metastable state,
we first displaced the \AsGa atom tentatively by 1.2~\AA\ 
from the lattice point to a puckered interstitial position 
along a bond breaking [111] direction, 
and then relaxed the lattice 
by means of the molecular dynamics with plane wave bases.
 The resultant distance after the relaxation
between \Asi and a neighboring As atom
on the axis of the symmetry \Ctv
was 1.37~\AA, 
a little larger than 
the previous values.~\cite{chadi1988, dabrowski1988, dabrowski1989}
The electron density were calculated 
for the perfect lattice of this size (64 atoms) and for lattices
in an extended size of $4 \times 2\sqrt{2} \times 3/\sqrt{2}$ (192 atoms) 
in which we embedded a relaxed $2 \times 2 \times 2$ supercell 
containing an \AsGa or an \Asi-\VGa\ pair.
 In these calculations, we fixed the atoms 
beyond the third neighbors at the perfect lattice positions.

%___________________________________________________
\begin{figure}
  \resizebox{55mm}{!}{\includegraphics{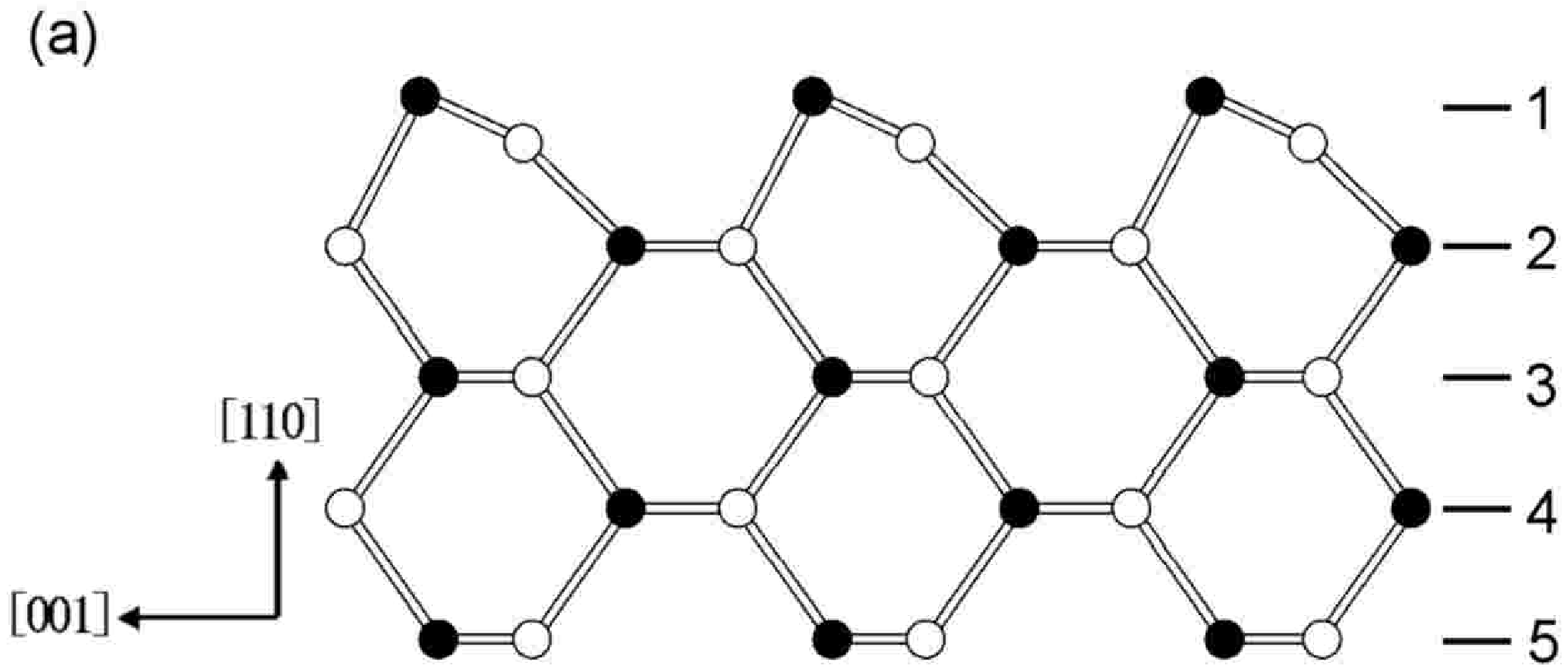}}
  \resizebox{20mm}{!}{\includegraphics{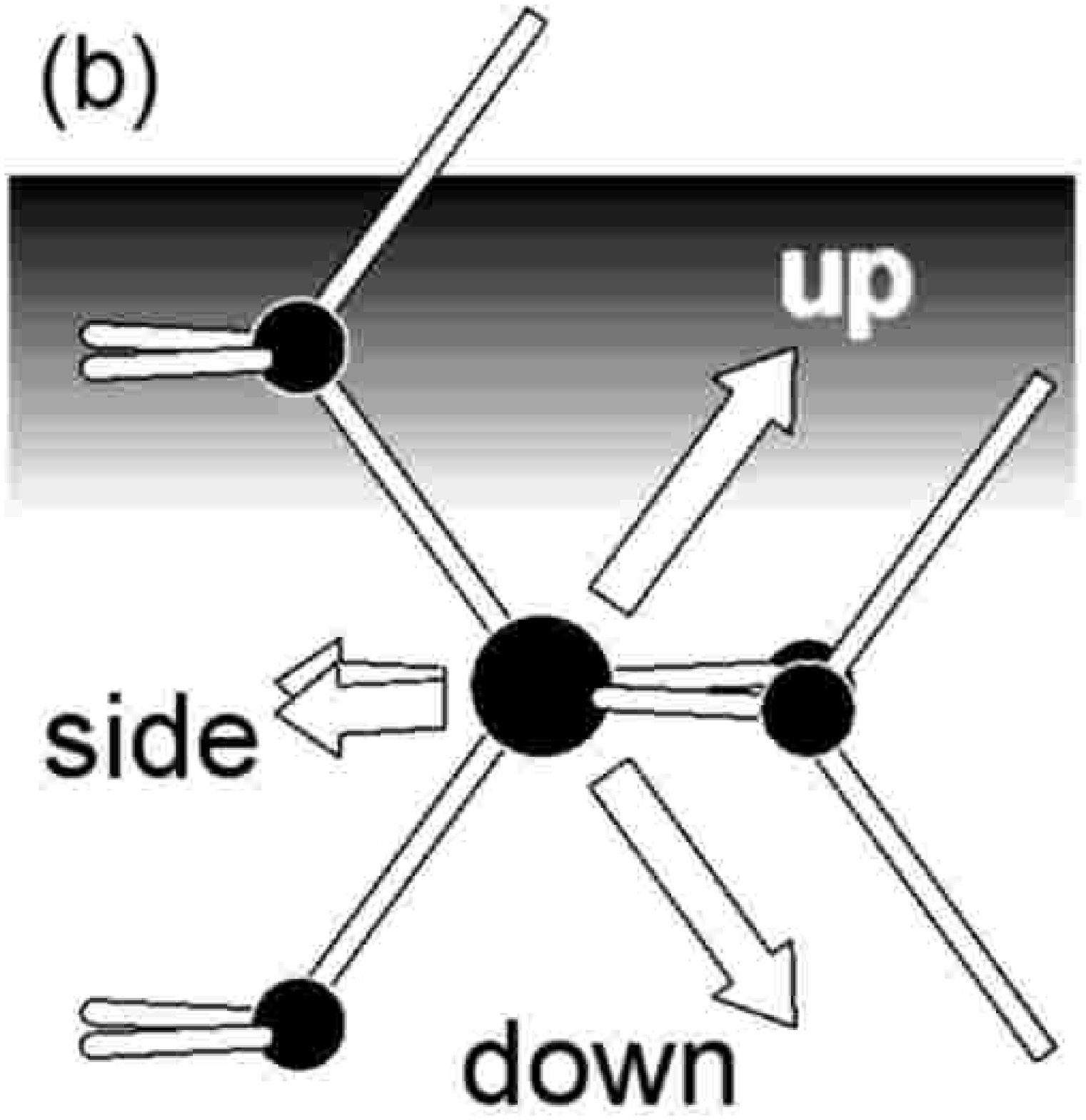}}
  \caption{\label{fig:def-layer}
 The definition of
(a) the layer number and 
(b) three inequivalent directions to which the \AsGa\ is displaced
to four possible metastable positions.
} 
\end{figure}

  For \AsGa defects near a (110) surface, 
we prepared a slab lattice parallel to the (110) plane 
in which one surface was relaxed 
and the other surface was terminated with hydrogen atoms.
 For convenience, we refer to the surface layer without hydrogen, 
though buckled as shown in Fig.~\ref{fig:def-layer}(a), 
as `layer 1', 
the one layer down as `layer 2', 
one more layer down as `layer 3' and so on.
 The electronic structures of the system bounded with surfaces 
were calculated for various configurations: 
without \AsGa defects, 
with an \AsGa defect (the stable state) 
located in the layer 1 to the layer 4.

 The procedure of lattice relaxation and 
electronic structure calculations are as follows.
 We first relaxed the surface bounded slab of 
a lateral size of $3 \times 3/\sqrt{2}$ and a thickness of 7 layers 
including the layer of hydrogen atoms.
The lattice constant was fixed in this case too.
 The positions of atoms 
in the relaxed configuration displaced 
from those of the ideal atomic positions 
are smaller than 0.2~\AA, 0.1~\AA\ and 0.05~\AA\ for the first, 
the second and the third neighbor atoms of \AsGa , respectively.
 Some of the neighbor atoms were displaced considerably, 
especially near the surface.
 For example,
for an \AsGa located on the layer 2, 
the third neighbor atoms on the layer 1 were displaced by 0.063~\AA, 
whereas the third neighbors on the layer 3 only by 0.019~\AA.
 Therefore, the atoms at shallow depths from the surface 
were not fixed tightly while the atoms further than third neighbors 
were fixed since all these atoms, even if relaxation was allowed, 
were found to be displaced by only a magnitude smaller than 0.07~\AA.
 After relaxing the small system, 
we embedded it in a larger system with a surface of a size of
 $4\times 4/\sqrt{2}$ and 5 layers including the hydrogen layer.
 For an \AsGa on the layer 4,
we embedded the relaxed system 
in a still larger system of the same surface size 
$4\times 4/\sqrt{2}$ but of 6 layers in thickness.

For calculations of metastable \Asi-\VGa\ pairs, 
we should note that there are three inequivalent configurations 
regarding the displacement direction 
relative to the surface as shown in Fig.~\ref{fig:def-layer}(b). 
We hereafter call the ``up'' configuration 
for the displacement direction ascending to the surface, 
the ``side'' configuration 
for the direction parallel to the surface, 
and the ``down'' configuration 
for the direction descending from the surface.
 The lattice relaxed to 
a metastable configuration similar to 
that in the bulk crystal except for \AsGa atoms on the layer 1,
where the similar displacement of the surface 
\AsGa atoms would destabilize the surface buckling 
and therefore is energetically unfavorable.

According to Tersoff-Hamann theory,~\cite{tersoff1985} 
filled-state STM images under a negative sample bias of $-2.3$~V, 
the bias condition experimentally employed 
in the previous study,~\cite{hida2001} 
were simulated by taking into account only the highest occupied state 
that should give the main contribution to the contrast.
 The image was calculated 
as a two-dimensional contour map of 
the iso-surface of electron density 
of the highest occupied state in vacuum out of the surface.

%___________________________________________________
\begin{figure*}[t]
  \resizebox{40mm}{!}{\includegraphics{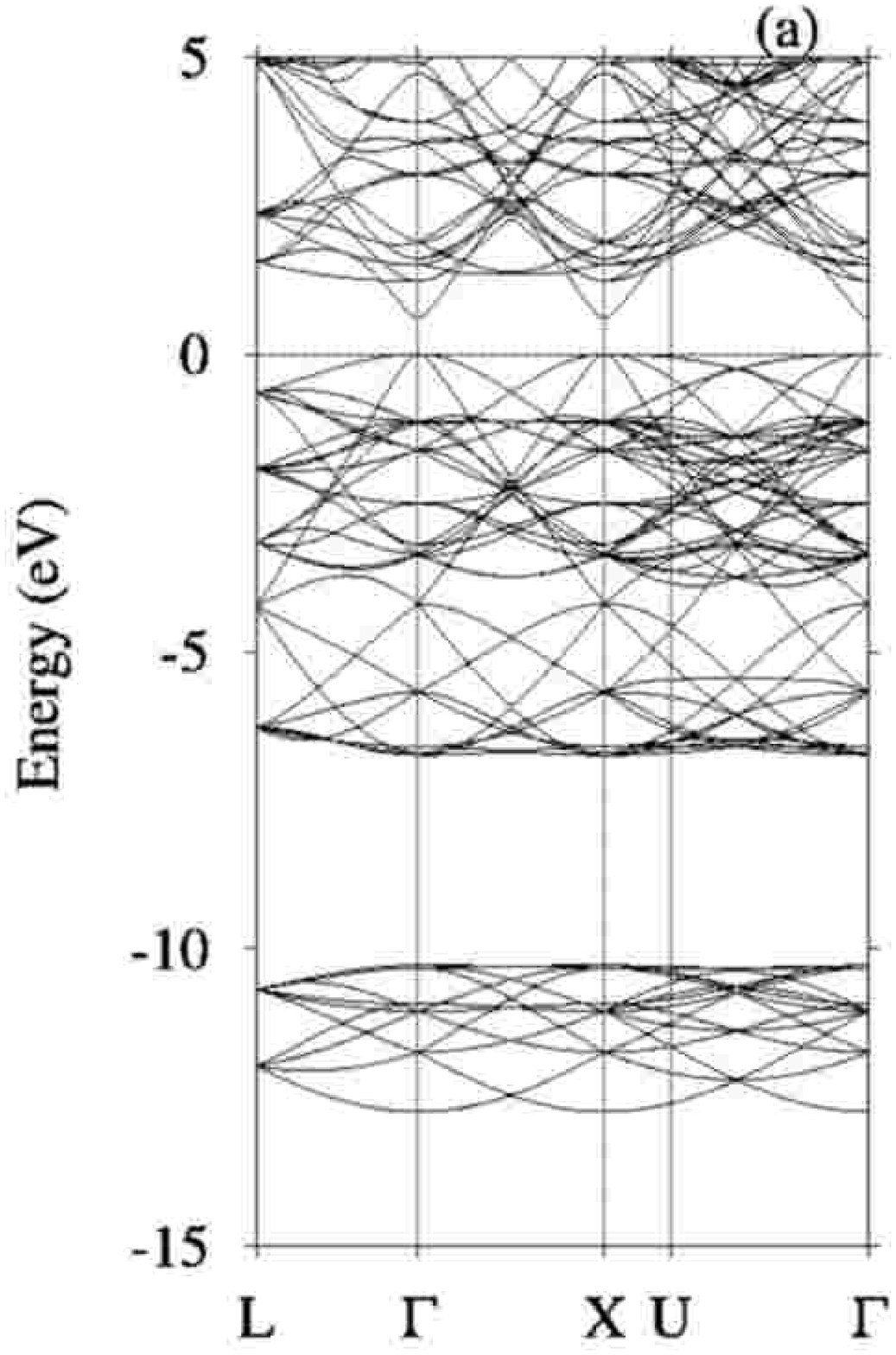}}
  \resizebox{40mm}{!}{\includegraphics{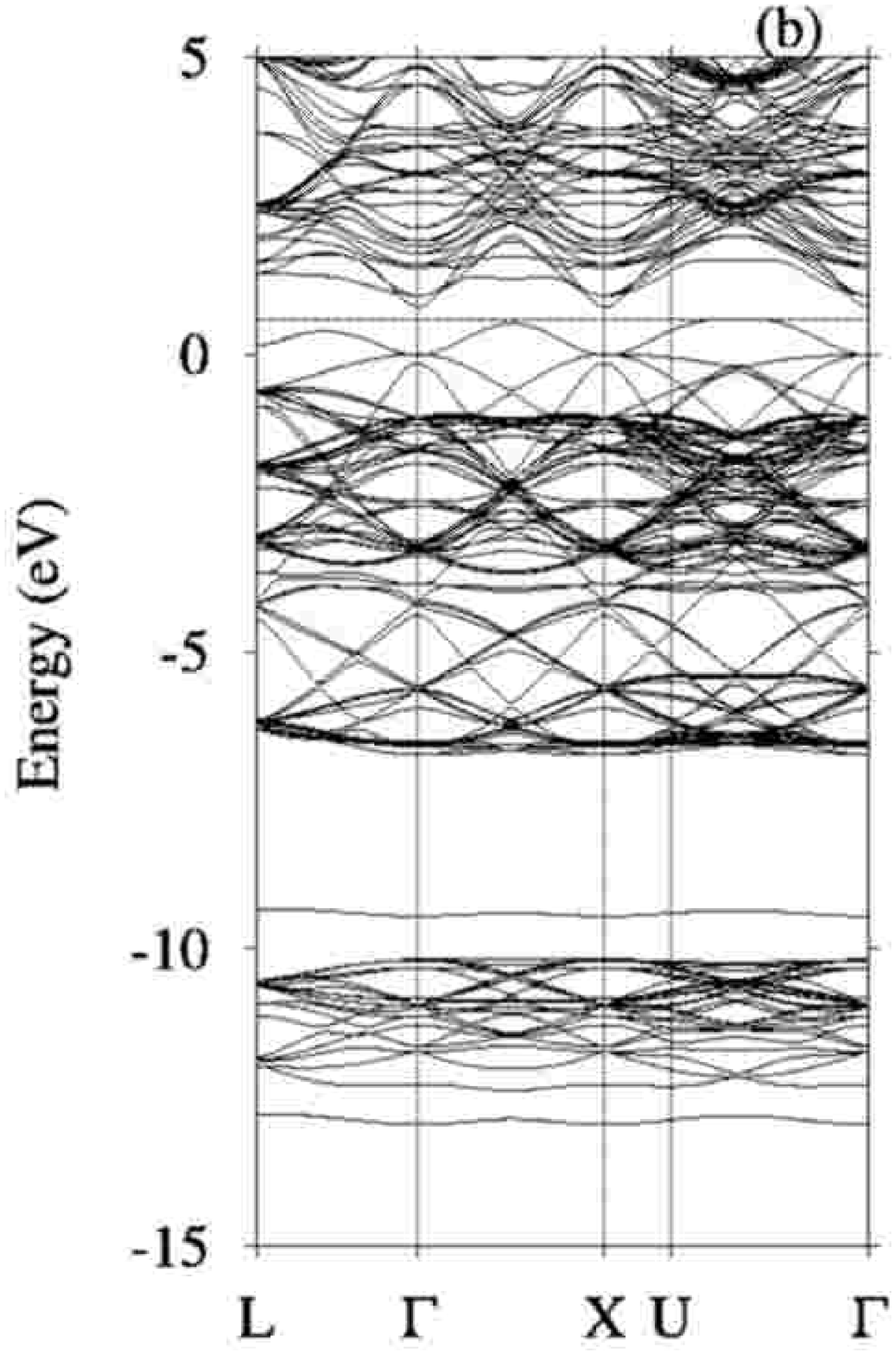}}
  \resizebox{40mm}{!}{\includegraphics{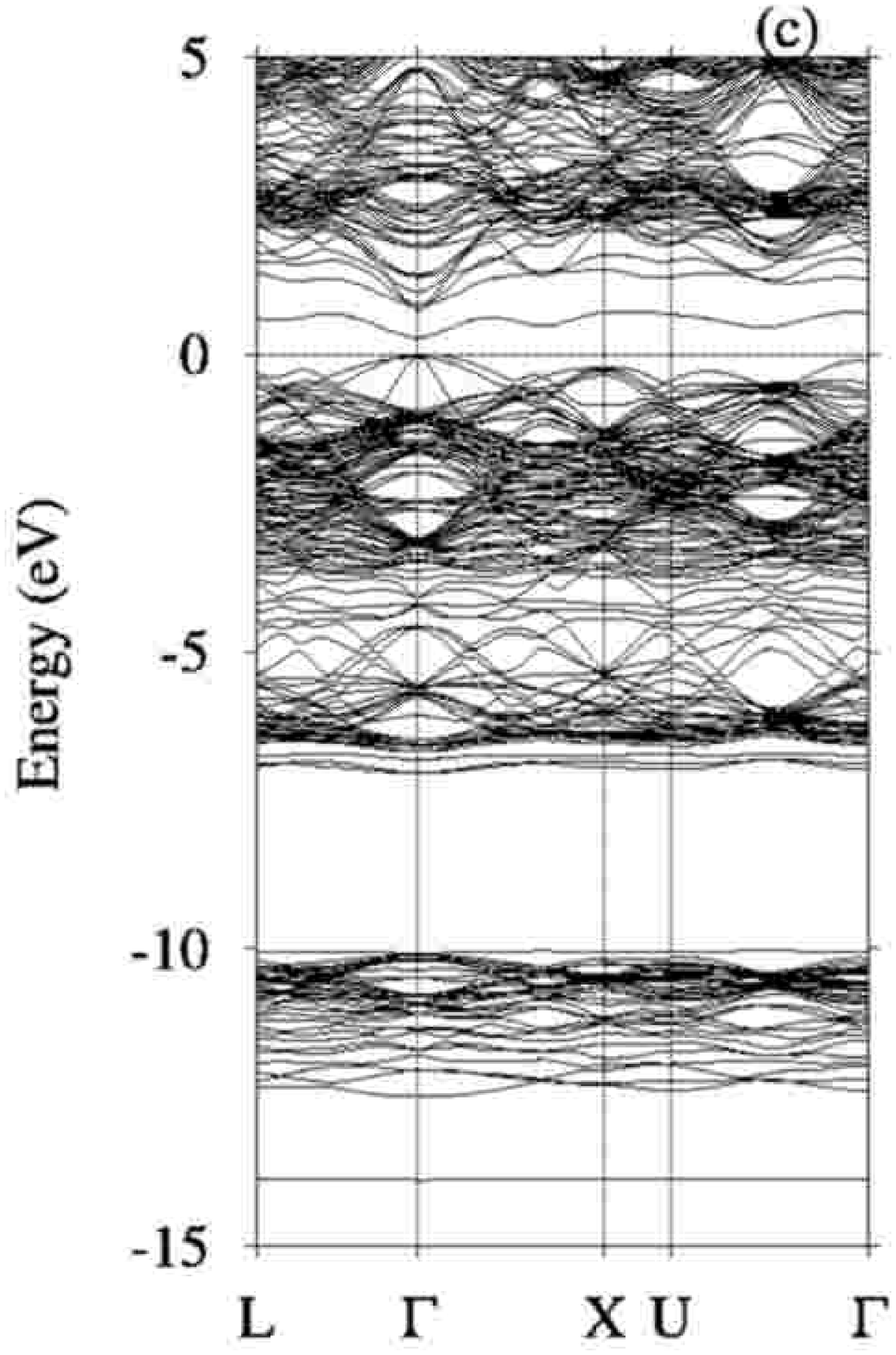}}\\
  \resizebox{40mm}{!}{\includegraphics{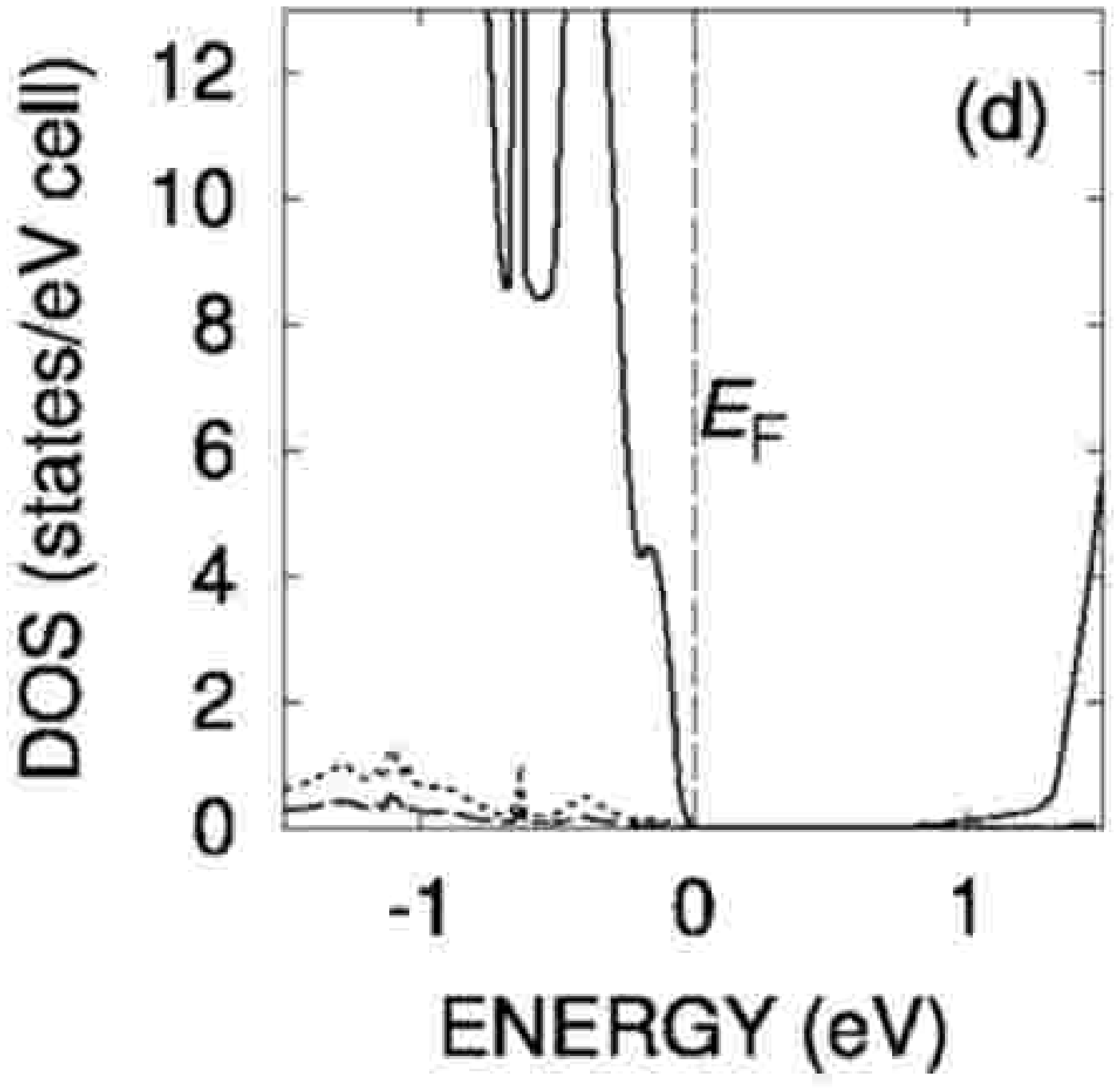}}
  \resizebox{40mm}{!}{\includegraphics{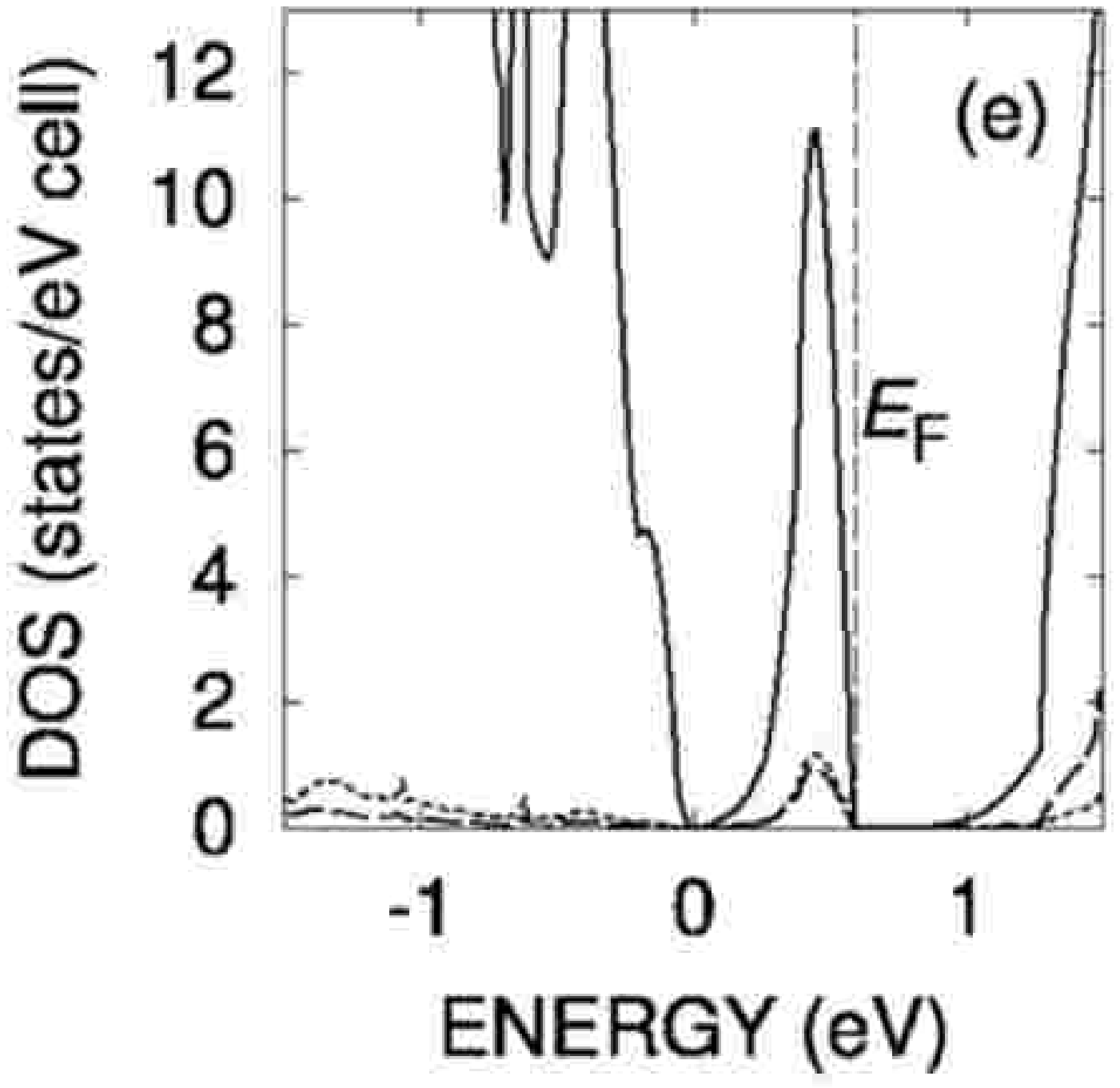}}
  \resizebox{40mm}{!}{\includegraphics{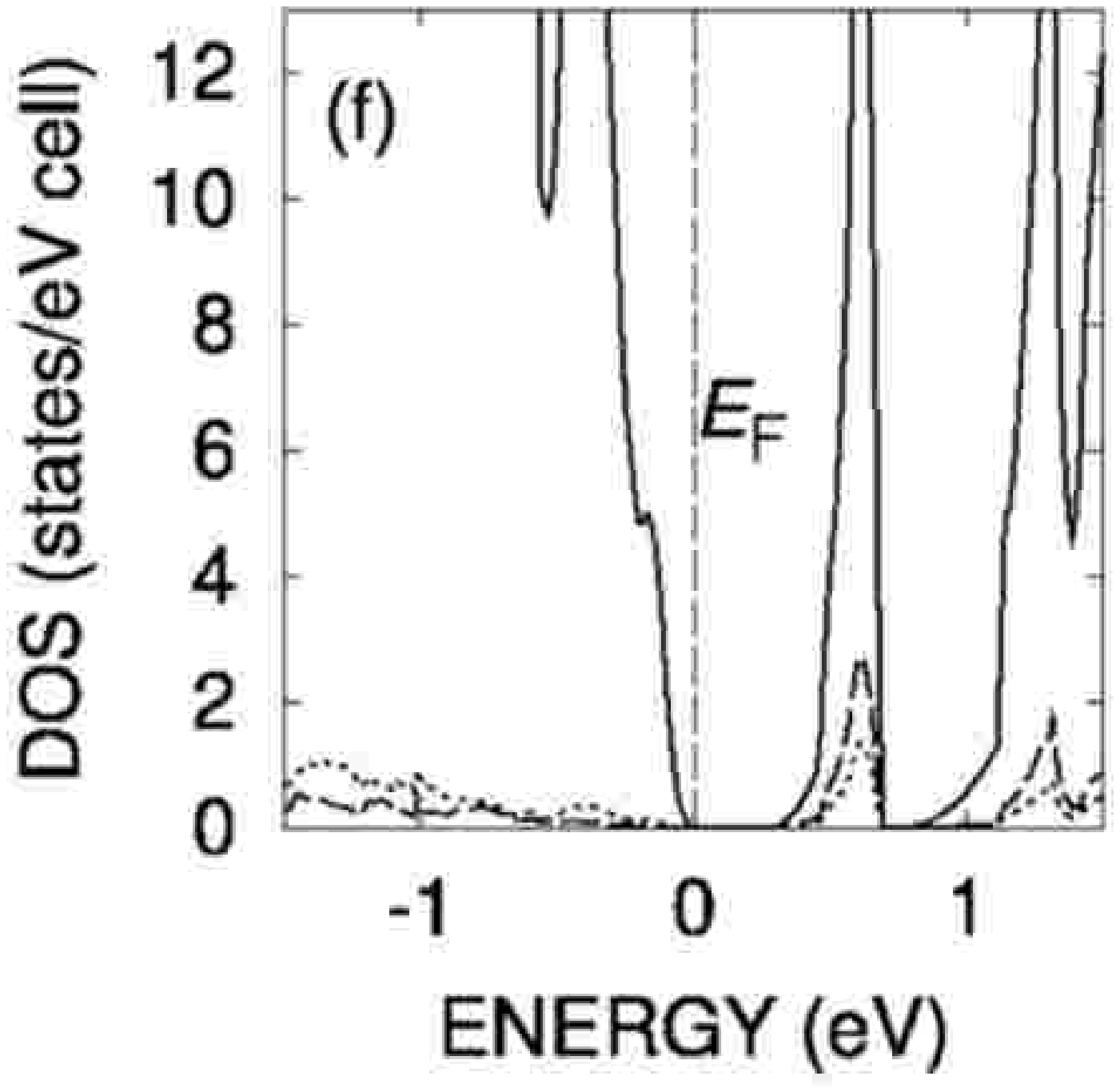}}\\
  \caption{\label{fig:bands} 
The band structure, 
the total density of states (DOS), 
and the local density of states (\lDOS) in the energy range of 
the band gap in a  relaxed system with a unit cell of 64 atoms. 
The energy zero is set to be at the position of the top of 
the bulk valence bands. 
  (a) The $E-{\bf k}$ relation without defect.
  (b) The $E-{\bf k}$ relation with \AsGa 
in the stable configuration.
  (c) The $E-{\bf k}$ relation with \Asi-\VGa pair
in the metastable configuration. 
  (d) The DOSs without defect corresponding to (a).
      The solid line: total DOS, 
      broken line: \lDOS\ of Ga, 
      dotted line: \lDOS\ of As.  
  (e) The DOSs with \AsGa corresponding to (b).
      The solid line: total DOS, 
      broken line: \lDOS\ of As$_{\rm Ga}$, 
      dotted line: averaged \lDOS\ of neighboring four As atoms. 
  (f) The DOSs with \Asi-\VGa pair corresponding to (c).
      The solid line: total DOS, 
      broken line: \lDOS\ of As$_{\rm i}$, 
      dotted line: averaged \lDOS\ of neighboring four As atoms.  
 In the $E-{\bf k}$ relations, the dashed horizontal lines represent
the Fermi energies.
 In comparison with the stable configuration (e), 
the Fermi energy $E_{\rm F}$ shifts
in the metastable configuration (f) 
and the gap state becomes unoccupied.
}
\end{figure*}

%___________________________________________________
\begin{figure}[b]
  \resizebox{60mm}{!}{\includegraphics{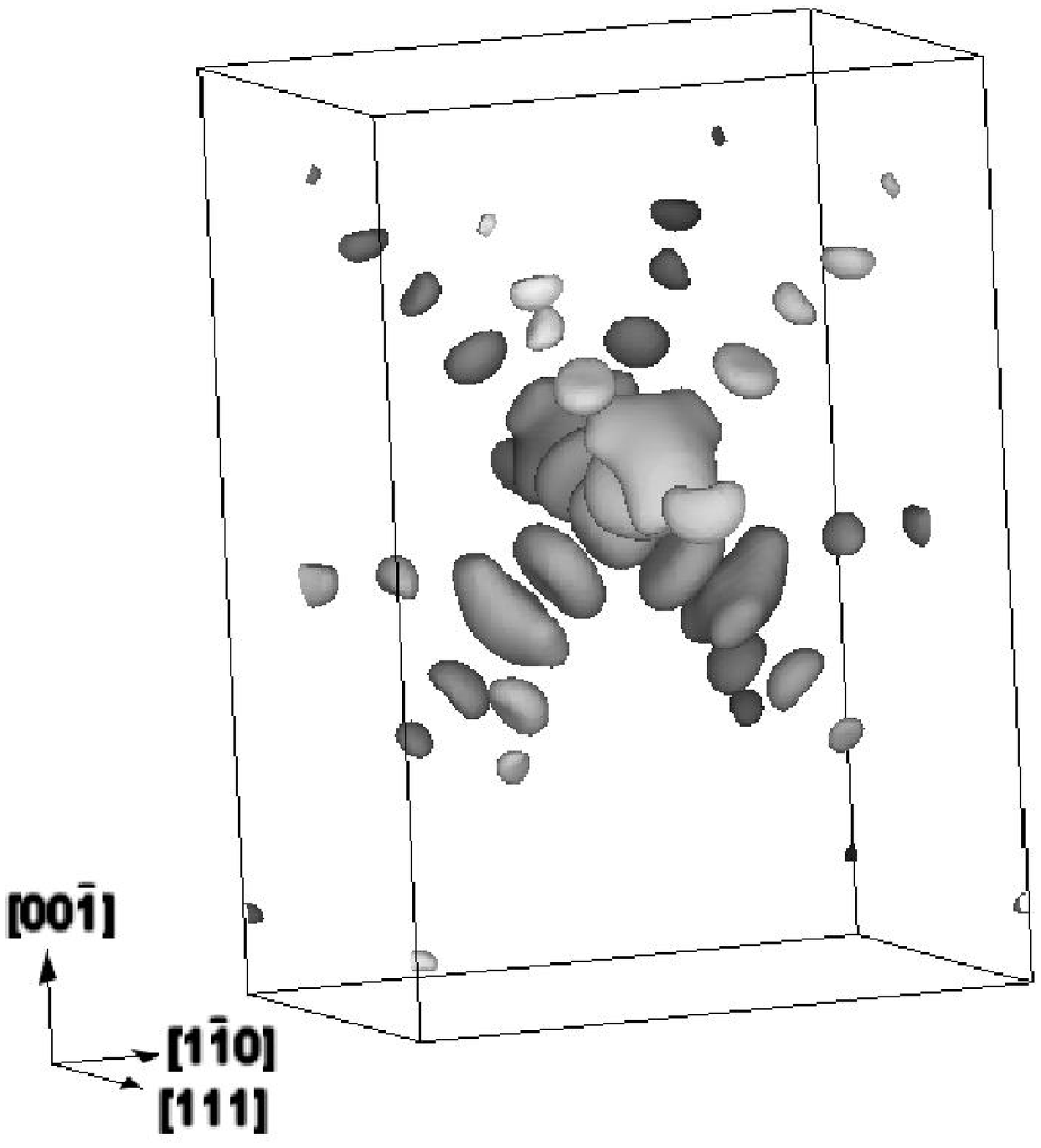}}\\
  \caption{\label{fig:bird-eye}
The typical iso-surface of the electron density of
the highest occupied band in the stable configuration in bulk.
}
\end{figure}
%___________________________________________________

%%%%%%%%%%%%%%%%%%%%%%%%%%%%%%%%%%%%%%%%%%%%%%%%%%%%%%%%%%%%%%%%%%%%%%%
%                    Electronic Structure                             %
%%%%%%%%%%%%%%%%%%%%%%%%%%%%%%%%%%%%%%%%%%%%%%%%%%%%%%%%%%%%%%%%%%%%%%%
\section{\label{sec:elec} Electronic Structure}

%=====================================================================%
%                   As antisite in bulk crystal                       %
%=====================================================================%
\subsection{As antisite in bulk crystal}

%---------------------------------------------------------------------%
%                        Stable configuration                         %
%---------------------------------------------------------------------%
\subsubsection{Stable configuration}

Figures \ref{fig:bands}(a) and \ref{fig:bands}(b) show
the $E-{\bf k}$ relations calculated by the LMTO method 
in a bulk system with a unit cell of 64 atoms. 
Figure \ref{fig:bands}(a) is that of the perfect lattice 
and \ref{fig:bands}(b) 
that of the lattice containing an \AsGa atom 
in the stable configuration. 
The band structure of the perfect lattice 
is characterized by three bands with energy gaps, 
the As s band below $-10.0$~eV, 
the valence band 
between $0.0$~eV and $-7.5$~eV, 
and the conduction band above the energy gap.
 An \AsGa atom in the stable configuration 
introduces an occupied band with relatively small dispersion 
in the band gap 
as shown in Fig.~\ref{fig:bands}(b). 
The DOS of perfect lattice and that of the lattice 
containing an \AsGa in the stable configuration are
shown in Figs.~\ref{fig:bands}(d) and ~\ref{fig:bands}(e) respectively.
Comparison between the two figures
indicates that \AsGa atom introduces the gap state 
mainly localized at the \AsGa atom 
and  neighboring four As atoms.

Figure \ref{fig:bird-eye} 
shows the iso-surface of electron density 
of the highest occupied state in the gap calculated 
by the plane wave basis for a system of 192 atoms 
containing an \AsGa in the stable configuration. 
One may see the \Td\ ~symmetric structure extending 
first along four [111] directions 
and then each further branching out along three [110] directions. 
The decreasing electron density with distance 
from the \AsGa shows
the localized nature of the corresponding 
\lDOS\ at the \AsGa atom (Fig.~\ref{fig:bands}(e)).
The iso-surface of electron density is shown also in
Figs.~\ref{fig:iso-stable}(a) and ~\ref{fig:iso-stable}(b).
 We note that the electron density at the \AsGa site 
is s-like (colored in red) 
and those on the nearest neighbor As sites are p-like 
(colored in blue) with the lobes heading toward the \AsGa atom. 
These p-like iso-surface are respectively linked 
to the central s-like iso-surface with a distinct node. 
The electron density on the third nearest neighbor As atoms 
are also p-like (colored in green) with the lobes aligning 
in the same directions as those of the first neighbor As atoms 
and the fifth neighbors (colored in yellow) 
again forming nodes between them. 
Thus, the electronic structure 
of the highest occupied level 
in the stable configuration
has a radial pattern that are characterized 
by the mutually anti-bonding As p-orbitals.

The defect formation energy $\Omega$ of \AsGa 
in the unit cell of 64 atoms,
is written in As-rich limit
as  ~\cite{zhang1991}
\begin{equation}
  \Omega 
= 
 E_{\rm D} 
 - N_{\rm e} \mu_{\rm e} 
 - N_{\rm Ga} \mu_{\rm GaAs} 
 - (N_{\rm As} - N_{\rm Ga}) \mu_{\rm As(bulk)},
\end{equation}
where
$E_{\rm D}$ is the total energy 
of the system with \AsGa,
$\mu_e$ is the chemical potential of electron, 
$\mu_{\rm GaAs}$ is the energy per atomic pair of bulk GaAs, 
and $\mu_{\rm As(bulk)}$ is the energy per atom of pure bulk As.
$N_{\rm e}(=0)$ is the charge of the defect, 
and $N_{\rm Ga}$ and $N_{\rm As}$ are the numbers of atoms 
of each species in the system.
The defect formation energy is estimated to be 0.6 eV 
by using the $\Gamma$ point sampling 
for the total energy calculation.
However, this total energy is underestimated
since, as seen in the Fig.~\ref{fig:bands}(b),
$\Gamma$ point is the bottom of the dispersion of the midgap
impurity level.
Since the peak of the midgap impurity level locates at 0.31 eV 
above its bottom,
the amount of underestimated energy would be 0.6 eV.
By adding it to the total energy,
we estimate the defect formation energy to be 1.2 eV.
This value is consistent with those of the previous calculations 
1.4-1.8~eV.~\cite{zhang1991, schick2002}
%

%----------------------------------------------------------------------%
%                  Metastable Configuration                            %
%----------------------------------------------------------------------%
\subsubsection{Metastable configuration}

Figures \ref{fig:bands}(c) and \ref{fig:bands}(f) 
show the $E - {\bf k}$ relation, 
the total DOS  
and the \lDOS\ calculated 
by the LMTO method for a system with 
a unit cell of 64 atoms 
containing an \AsGa\ atom in the metastable configuration 
(\Asi-\VGa) 
that was fully relaxed by the pseudopotential method. 
The analysis of the \lDOS\ shows 
that the basis function localized at the \Asi atom constitutes 
the main contribution to the lowest unoccupied states 
introduced in the band gap 
between the valence band and the conduction band. 
Instead, there are no occupied states 
associated with localized orbitals 
at the \Asi atom near the Fermi level.
This implies that the lowest unoccupied states 
and the highest occupied states in the stable configuration 
exchange their energetic order in the metastable configuration. 
However, the position of the unoccupied  level in the gap 
was found to be sensitive to the computational detail.
For instance, a pseudopotential calculation with a larger supercell
shifts this unoccupied level  
to a slightly higher energy position
in the conduction band region.

\begin{figure*}
  \resizebox{60mm}{!}{\includegraphics{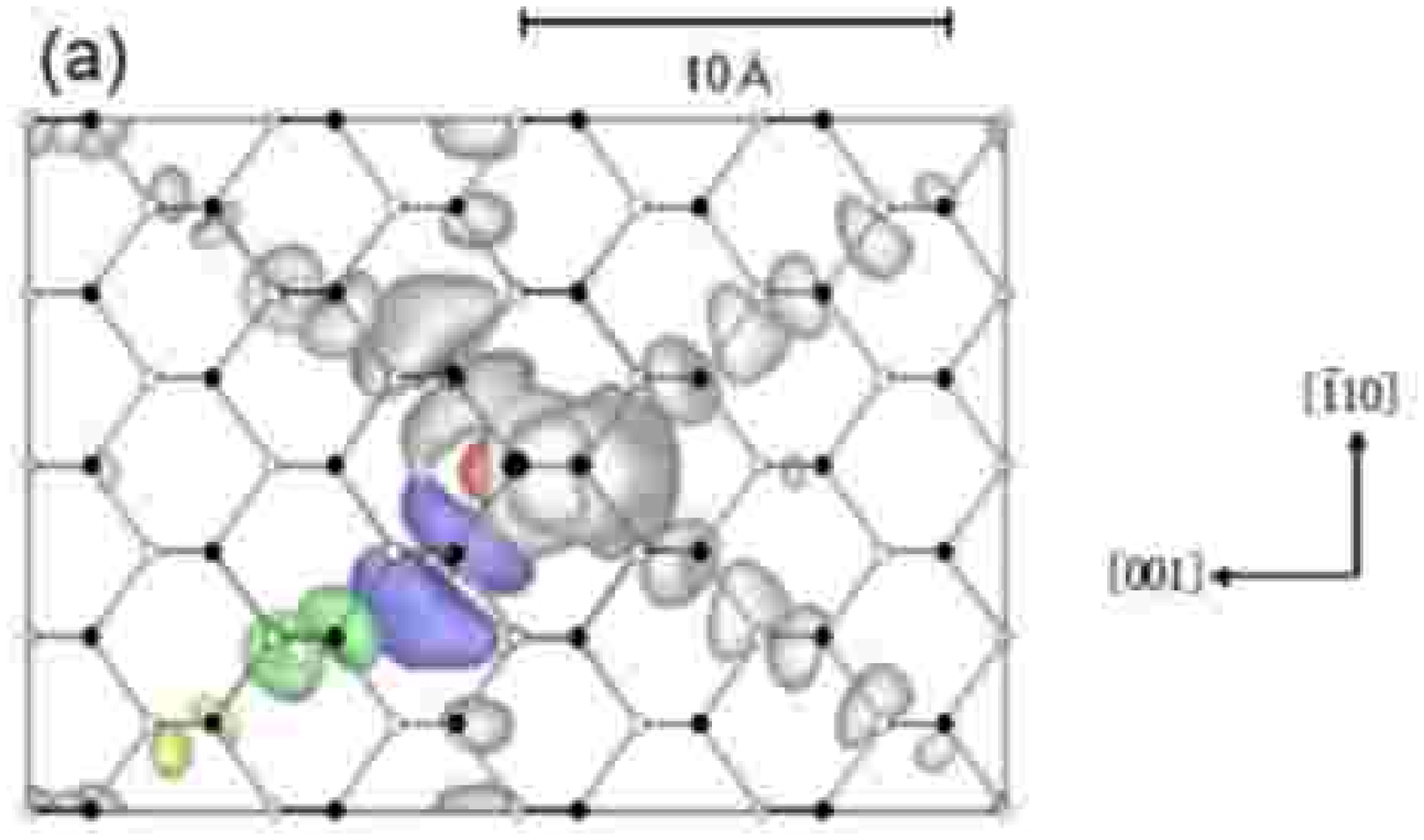}}
  \resizebox{60mm}{!}{\includegraphics{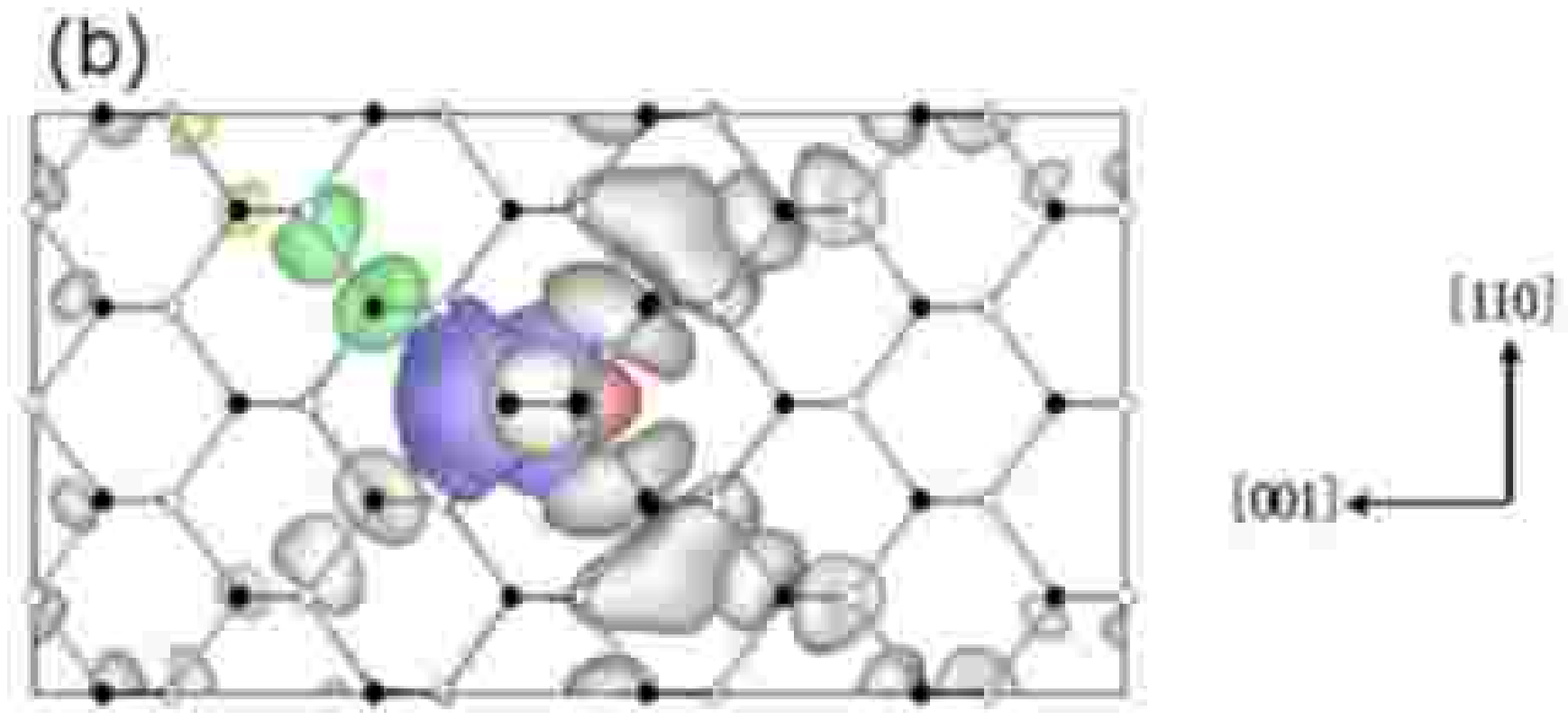}}\\
  \resizebox{60mm}{!}{\includegraphics{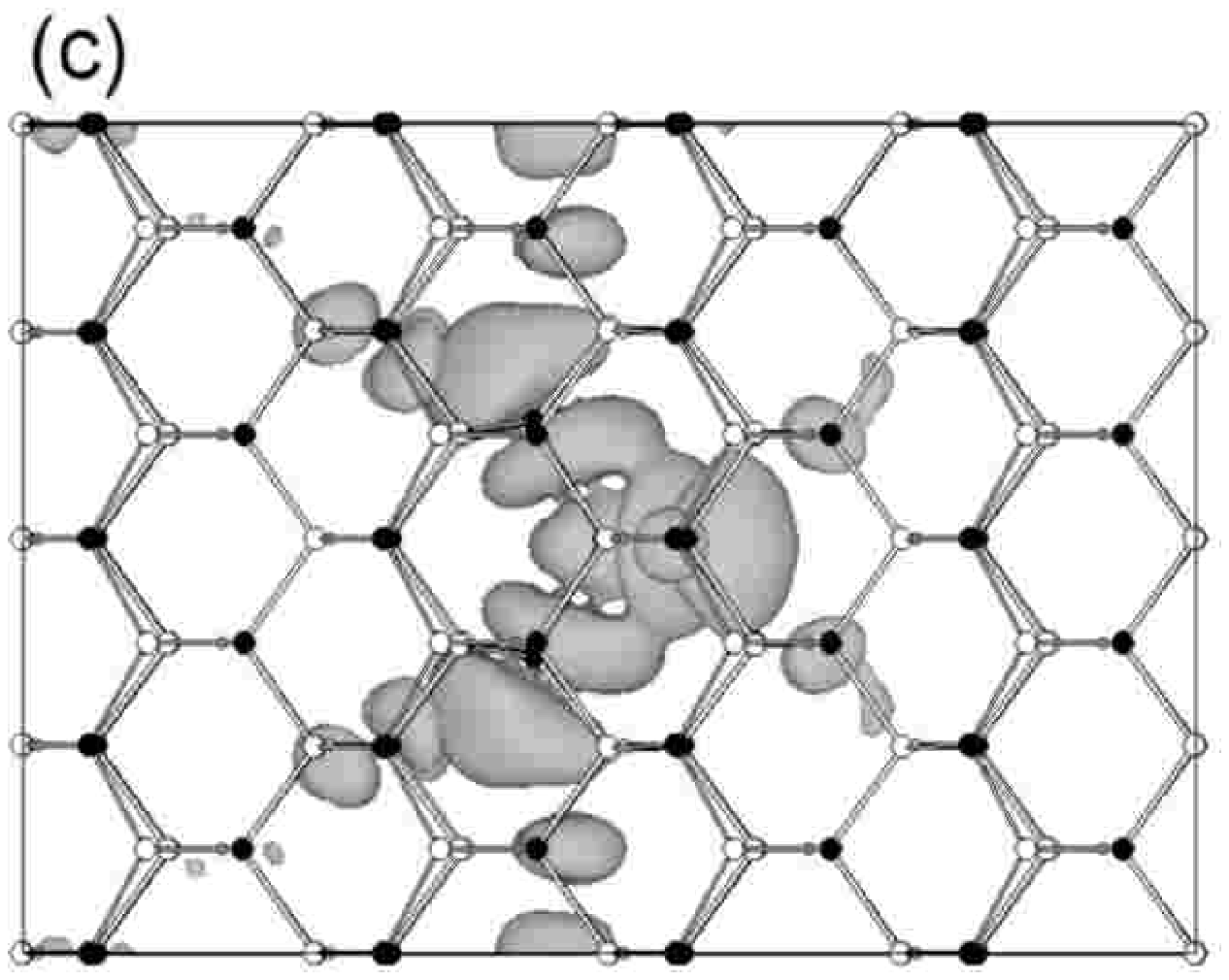}}
  \resizebox{60mm}{!}{\includegraphics{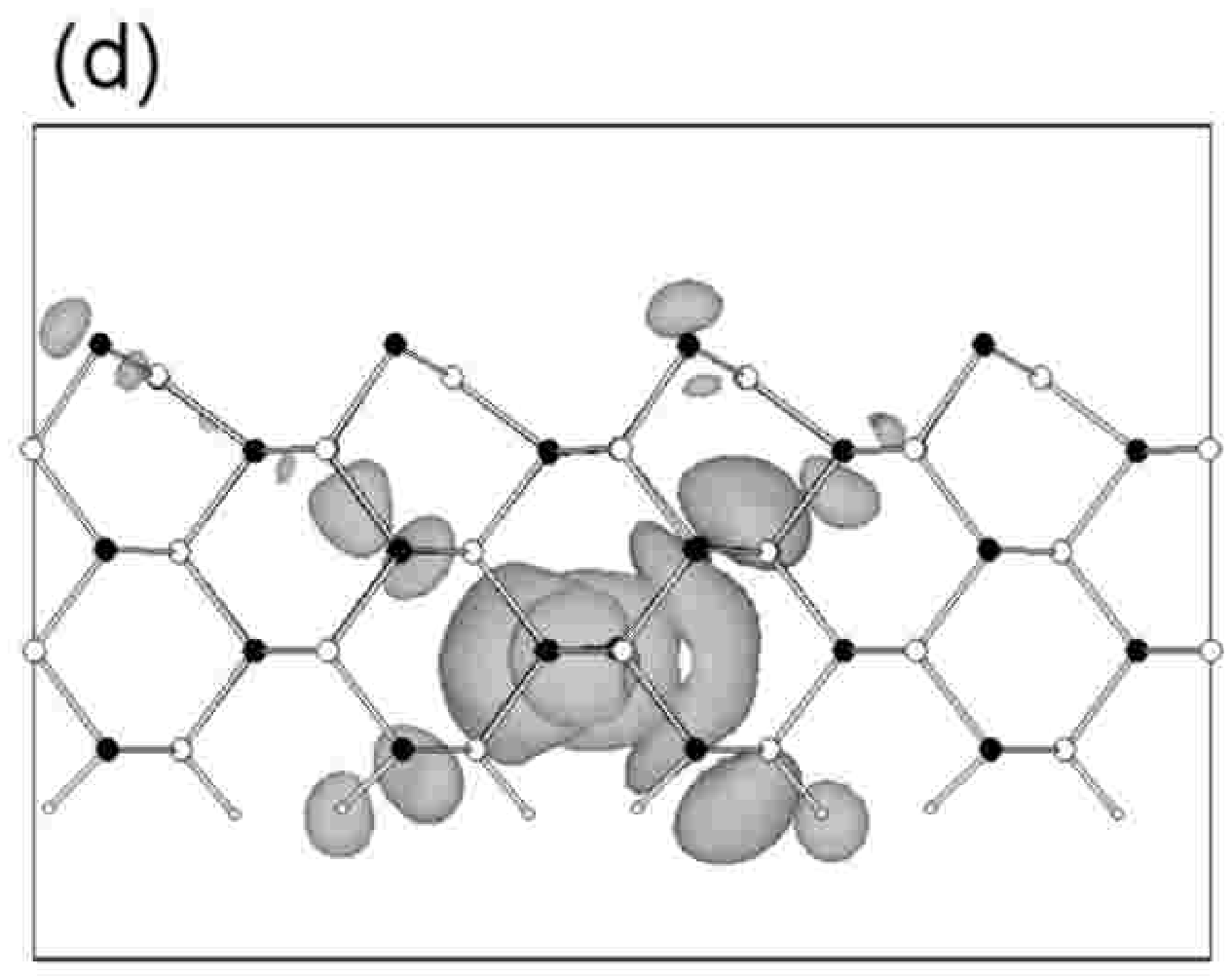}}\\
  \resizebox{60mm}{!}{\includegraphics{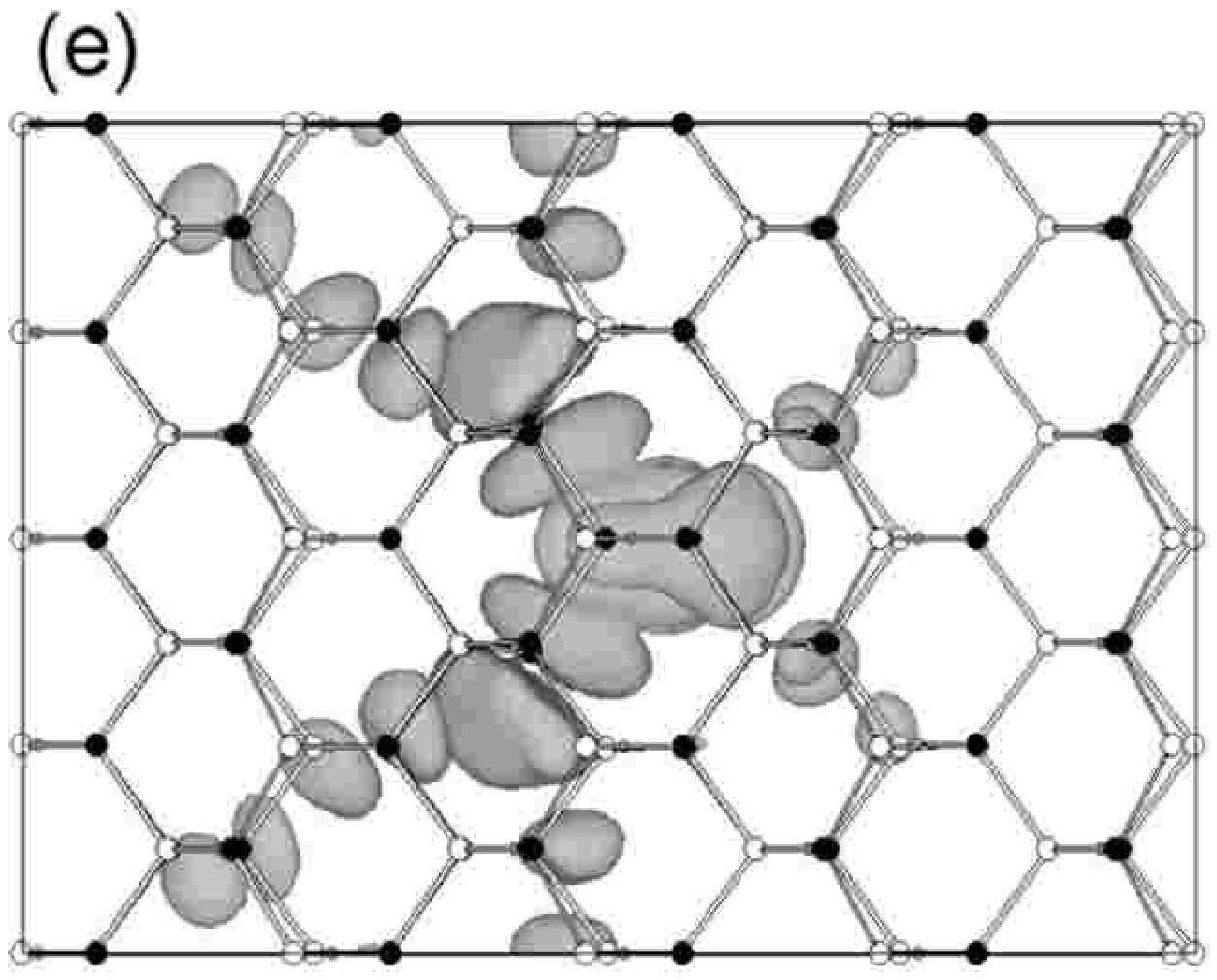}}
  \resizebox{60mm}{!}{\includegraphics{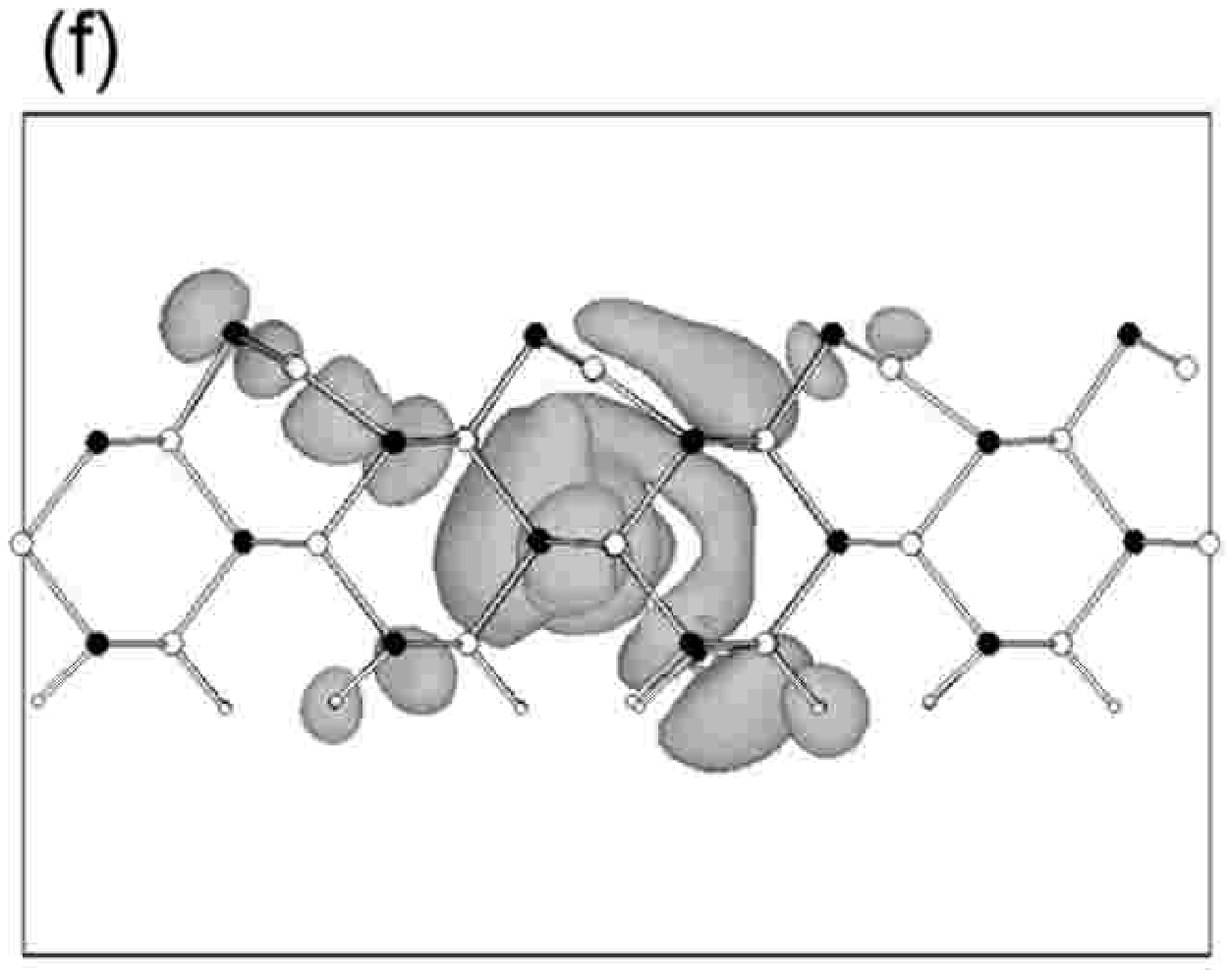}}\\
  \resizebox{60mm}{!}{\includegraphics{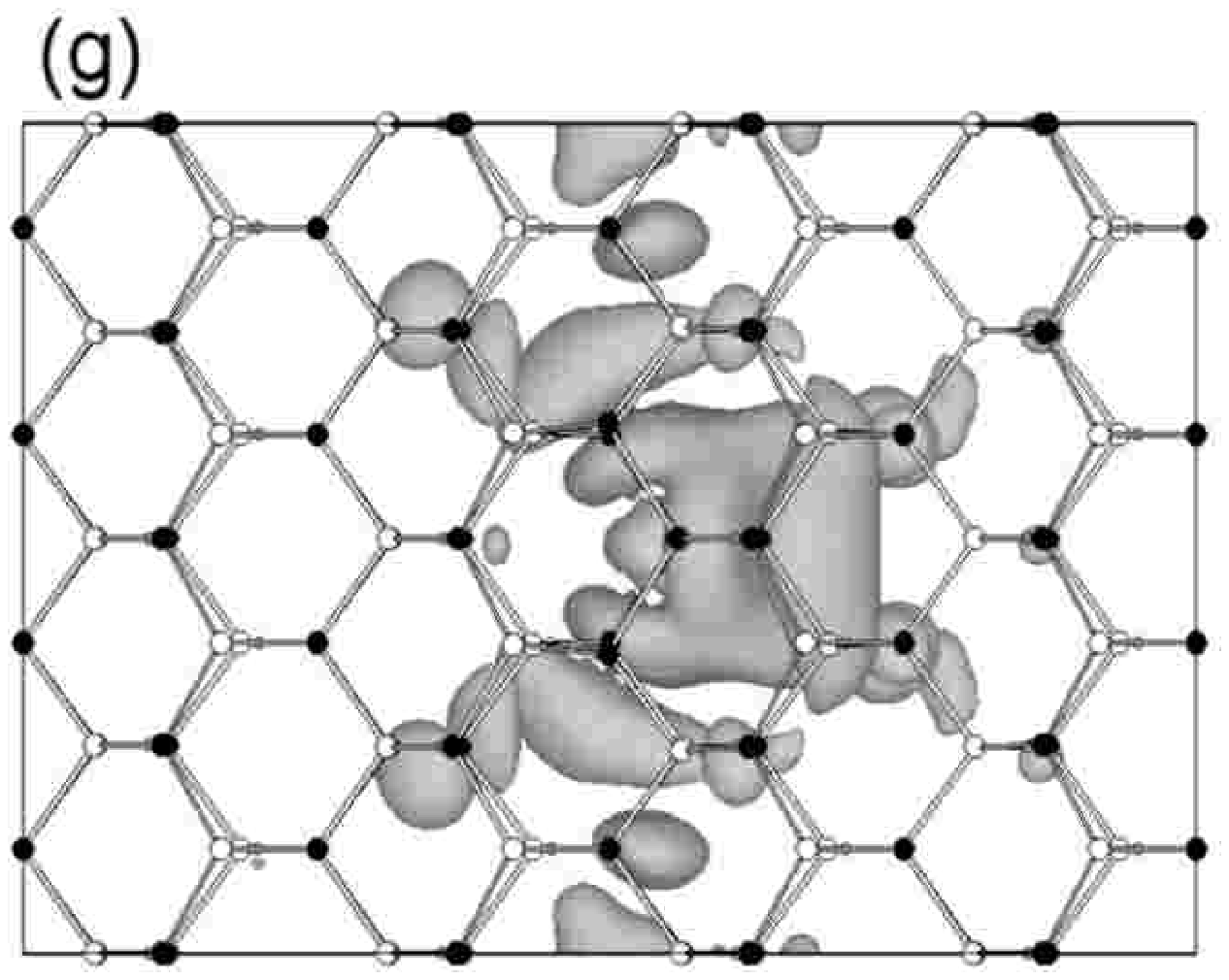}}
  \resizebox{60mm}{!}{\includegraphics{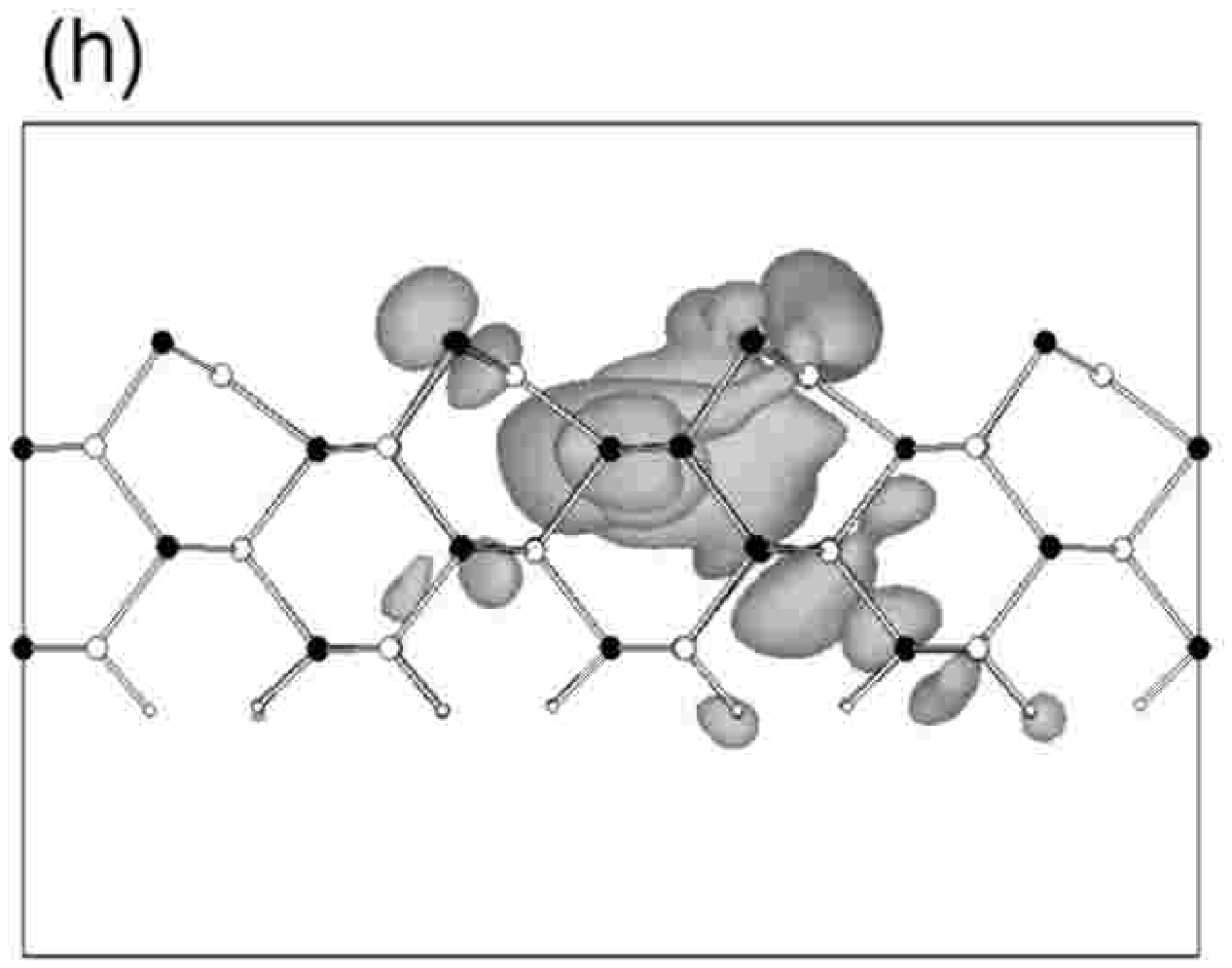}}\\
  \resizebox{60mm}{!}{\includegraphics{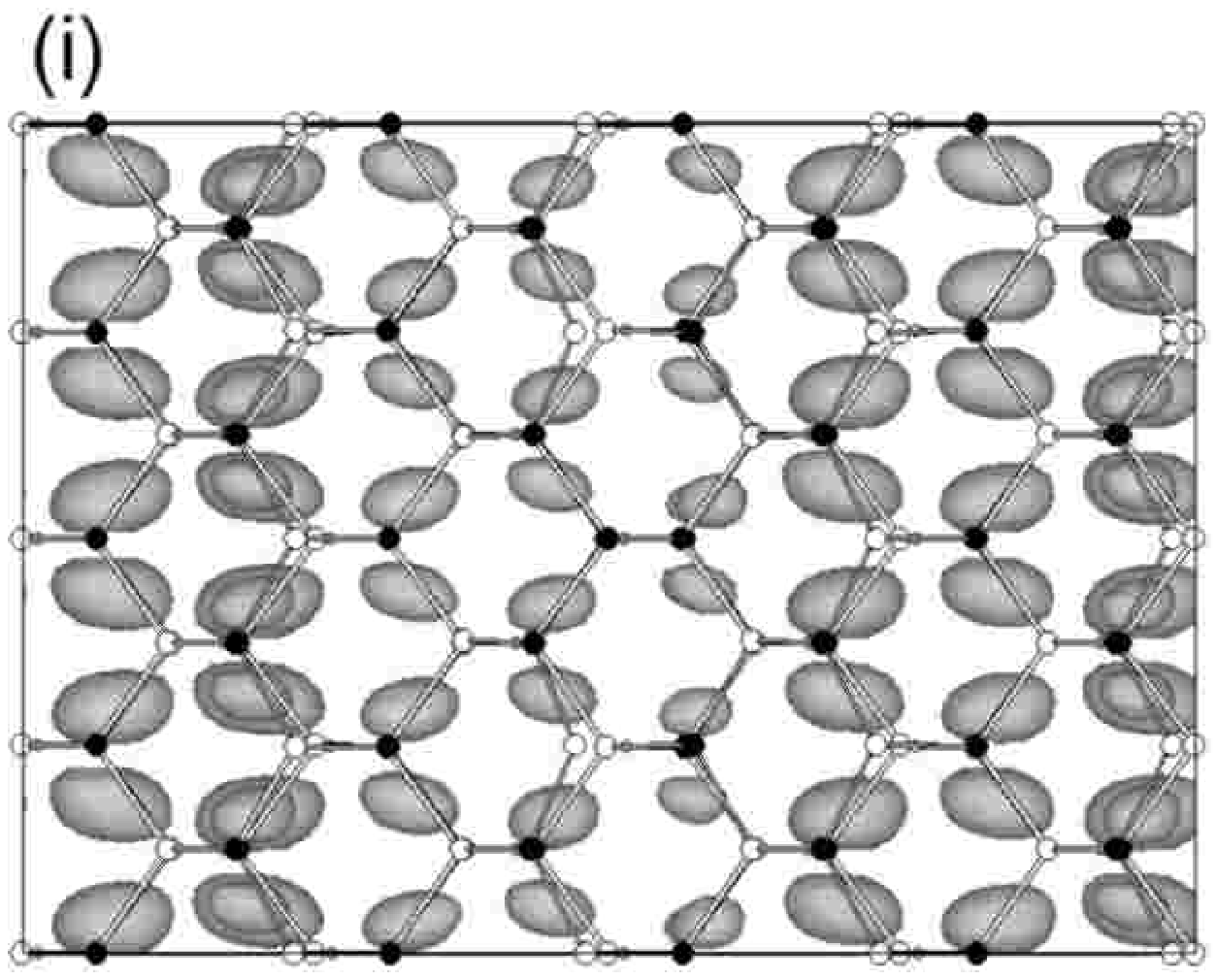}}
  \resizebox{60mm}{!}{\includegraphics{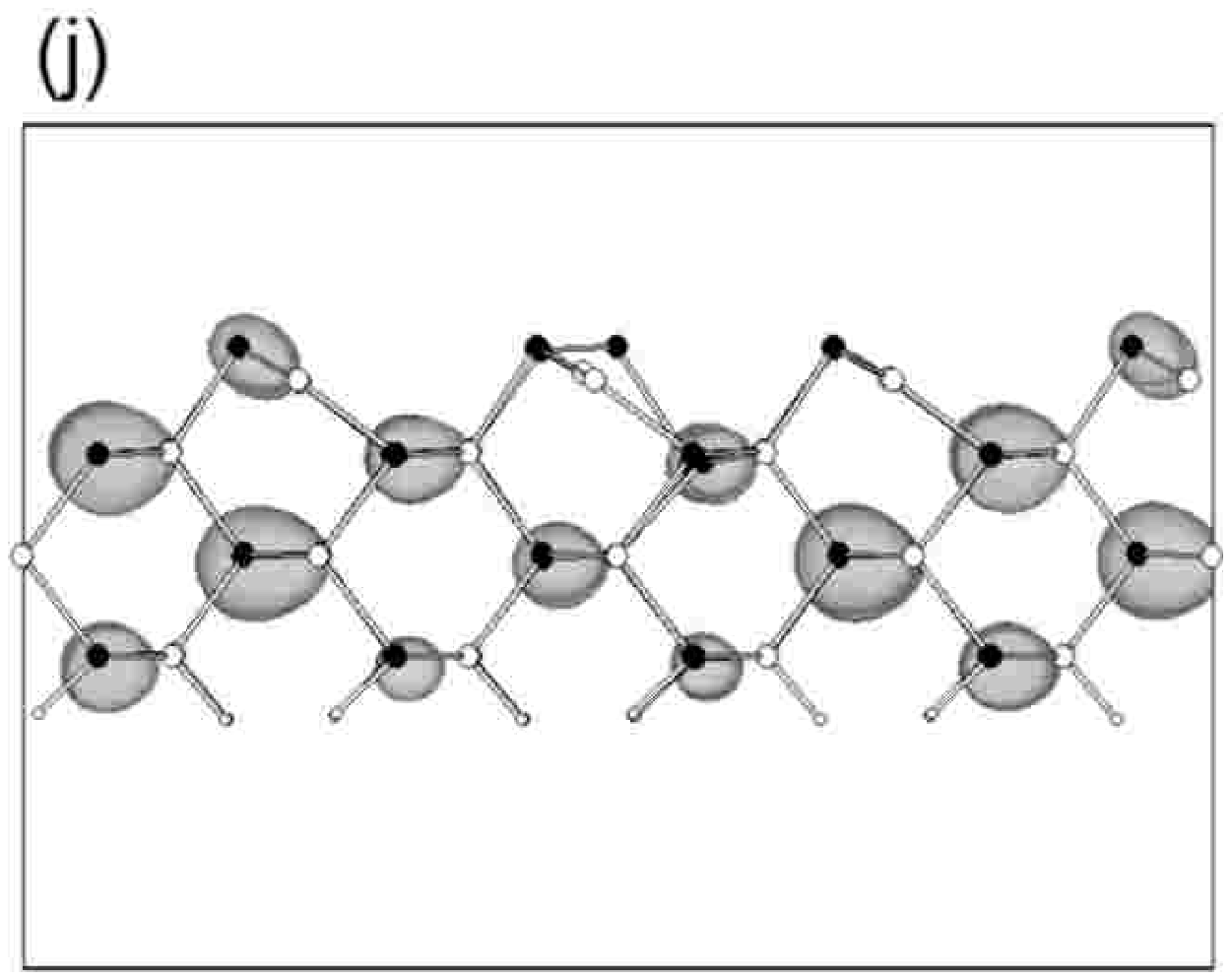}}\\
  \caption{\label{fig:iso-stable}
The iso-surface of the electron density 
of the highest occupied level
of \AsGa\ in the stable configuration
calculated by pseudopotential method.
The figures in left and right column are the top 
and side view of the electronic density, respectively.
Figures (a) and (b) represent the iso-surfaces 
of electron density $1.37 \times 10^{-3}$ e/\AA${}^3$\ in 
the bulk crystal.
Black spheres represent As atoms, and white spheres Ga atoms.
The color red, blue, green and yellow, 
represent the wavefunction on the central \AsGa, 
the first, the third  and the fifth neighbor 
As atoms, respectively. 
The wavefunction has little amplitude on the Ga site. 
The average of the electron density of this level is 
$2.31 \times 10^{-4}$ e/\AA${}^3$. 
In the stable configuration,
there are nodes 
between the s-orbital of the \AsGa (red) and 
p-orbitals of neighboring four As atoms. 
Figures (c) to (j) show the iso-surface of electron density
$2.26\times 10^{-3}$ e/\AA${}^3$\ for
an \AsGa\ in the layer 4 ((c), (d)), 
the layer 3 ((e), (f)), the layer 2 ((g), (h)),
and the layer 1 ((i), (j)).
The average electron density of this level is 
$1.73\times 10^{-4}$ e/\AA${}^3$.
}
\end{figure*}

Figures \ref{fig:iso-meta}(a) and \ref{fig:iso-meta}(b) show 
the electron density of the corresponding 
highest occupied state of the system of 192 atoms 
containing a relaxed \Asi -\VGa\ pair. 
The electron density of this state
extends over a wider range and differs substantially 
from the symmetric radial pattern in the stable configuration shown 
in Figs.~\ref{fig:iso-stable}(a) and \ref{fig:iso-stable}(b). 
The less localized feature is consistent 
with the large dispersion of the highest occupied band 
that merges into the bulk valence bands as seen 
in Fig.~\ref{fig:bands}(c). 
We should note that the iso-surface 
of the electron density at the \Asi atom 
becomes p-like in the metastable configuration 
in contrast to the s-like feature 
in the stable configuration. 
The p-like iso-surface at the \Asi is in parallel to those of 
the three neighboring As atoms with no nodes between them. 
The radial pattern characteristics 
of the stable configuration
is broken to \Ctv.
Among the iso-surface branches extending in four [111] directions,
the branch remained is only the one in the direction 
opposite to the displacement direction of the \AsGa atom.
The lobe of the branch, however, preserves 
the p-like anti-bonding character as in the stable configuration.

%======================================================================%
%                 As antisite located near (110) surface               %
%======================================================================%
\subsection{As antisite located near (110) surface}

%--------------------------------------------------------------
%--------------------------------------------------------------

%----------------------------------------------------------------------%
%                       Stable configuration                           %
%----------------------------------------------------------------------%
\subsubsection{Stable configuration}
For slab systems with a surface, 
the electronic energy levels were calculated 
only by the pseudopotential method with plane wave basis. 
Except for the \AsGa located in the surface layer (the layer 1), 
the highest occupied state remains amid 
in the band gap as in the bulk crystal. 
The absence of the gap state associated with \AsGa located 
in the layer 1 is shown also by the \lDOS\ 
calculated by the LMTO for a system of 60 atoms 
with the surface size of $2 \times 3/\sqrt{2}$
and the thickness of 5 layers 
terminated with hydrogens on one side. 
Figures 
\ref{fig:iso-stable}(c)-(j)
show the iso-surfaces of electron density 
of the highest occupied states in the gap 
for an \AsGa located in the stable configuration  
in the layer 1-4.
In all the cases, the features of the iso-surface is essentially 
the same as those in the bulk crystal 
(Figs. \ref{fig:iso-stable}(a) and \ref{fig:iso-stable}(b)): 
the s-like feature at the \AsGa site and the p-like ones 
at the neighboring As atoms.
The radial pattern characteristics of the highest occupied state 
in the bulk crystal look as if it were just terminated 
with the surface. 
Even in the \AsGa located in the layer 2
(Figs. \ref{fig:iso-stable}(g) and \ref{fig:iso-stable}(h)), 
though the electron density is concentrated mainly on the first 
and the third neighbor As atoms near the surface 
with more distinct p-like features than in the bulk lattice, 
the atomic configuration around the \AsGa almost remains the same 
as in the bulk lattice. 
The \AsGa located in the layer 3 
(Figs. \ref{fig:iso-stable}(e) and \ref{fig:iso-stable}(f)), 
however, shows a subtle difference 
in the iso-surface from that in the bulk crystal. 
The p-like iso-surface at the first nearest As atom 
extends considerably 
to Ga atoms in the layer 1 
forming $\pi$-like bonds without a node. 
This may be an effect of the surface buckling: 
For the surface Ga, As in the layer 2, and \AsGa, 
the angle of Ga-As-\AsGa is at 87$^\circ$, 
which is approximately right angle,
so that the p-orbitals on the As atoms can easily form $\pi$ bonds.
The situation is quite different 
when the As antisite is located in the layer 1, 
for which Figs. \ref{fig:iso-stable}(i) and \ref{fig:iso-stable}(j) show 
the iso-surface of the electron density. 
The cause of such a surface effect 
on the electronic structure is discussed
in Sec.~\ref{sec:discussion}.

%----------------------------------------------------------------------%
%                    Metastable configuration                          %
%----------------------------------------------------------------------%

%___________________________________________________
\begin{figure*}
  \resizebox{60mm}{!}{\includegraphics{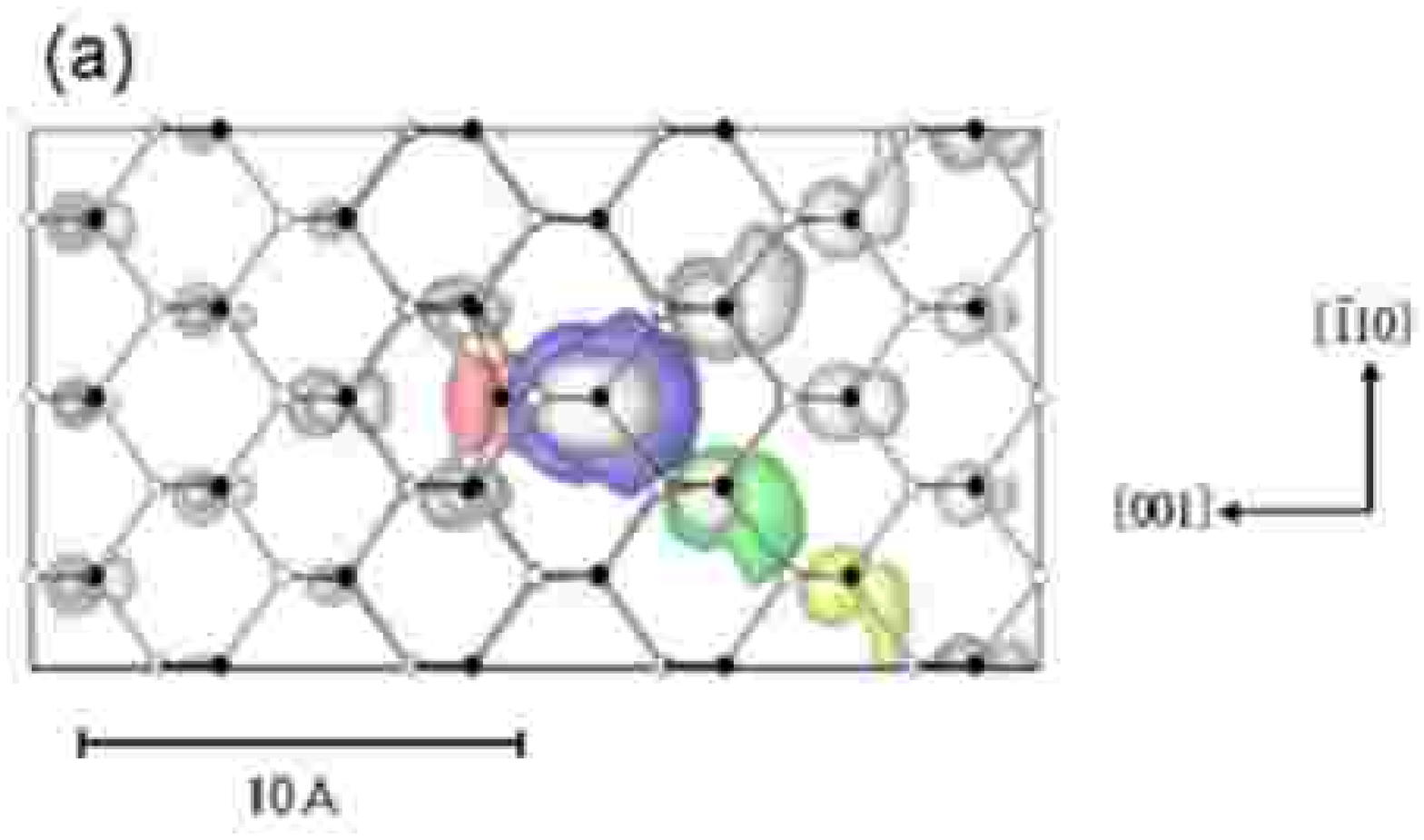}}
  \resizebox{60mm}{!}{\includegraphics{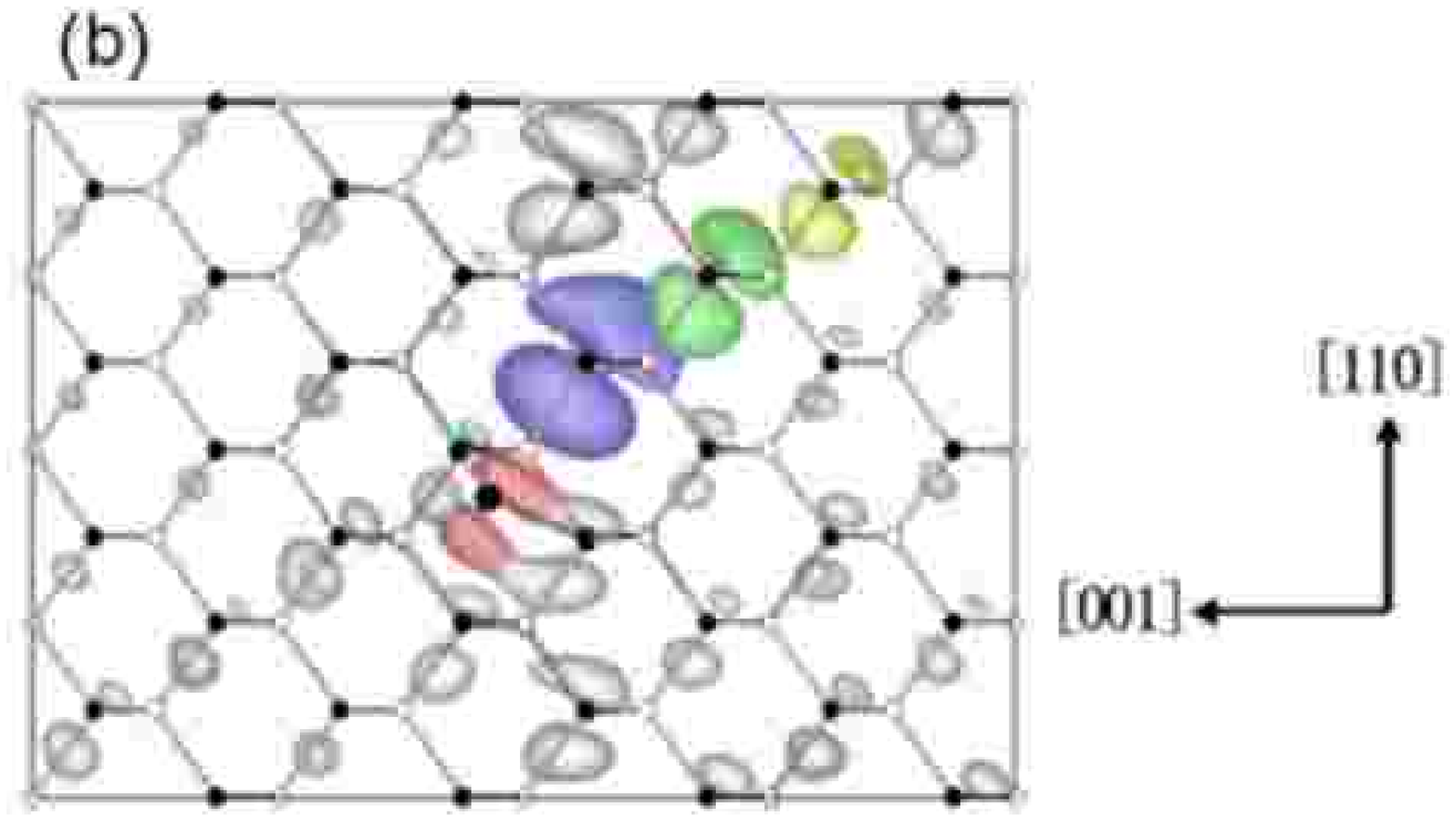}}\\
  \resizebox{60mm}{!}{\includegraphics{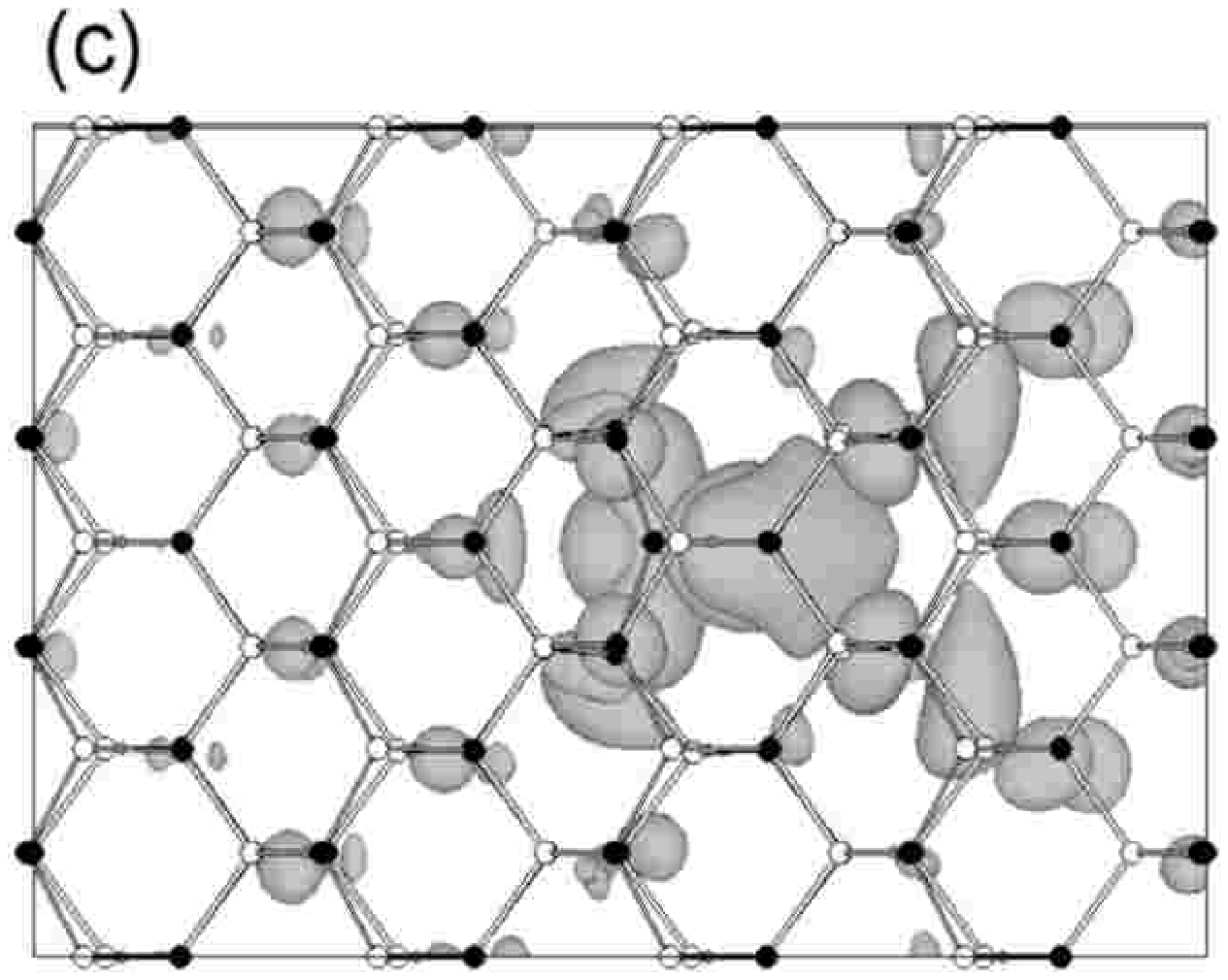}}
  \resizebox{60mm}{!}{\includegraphics{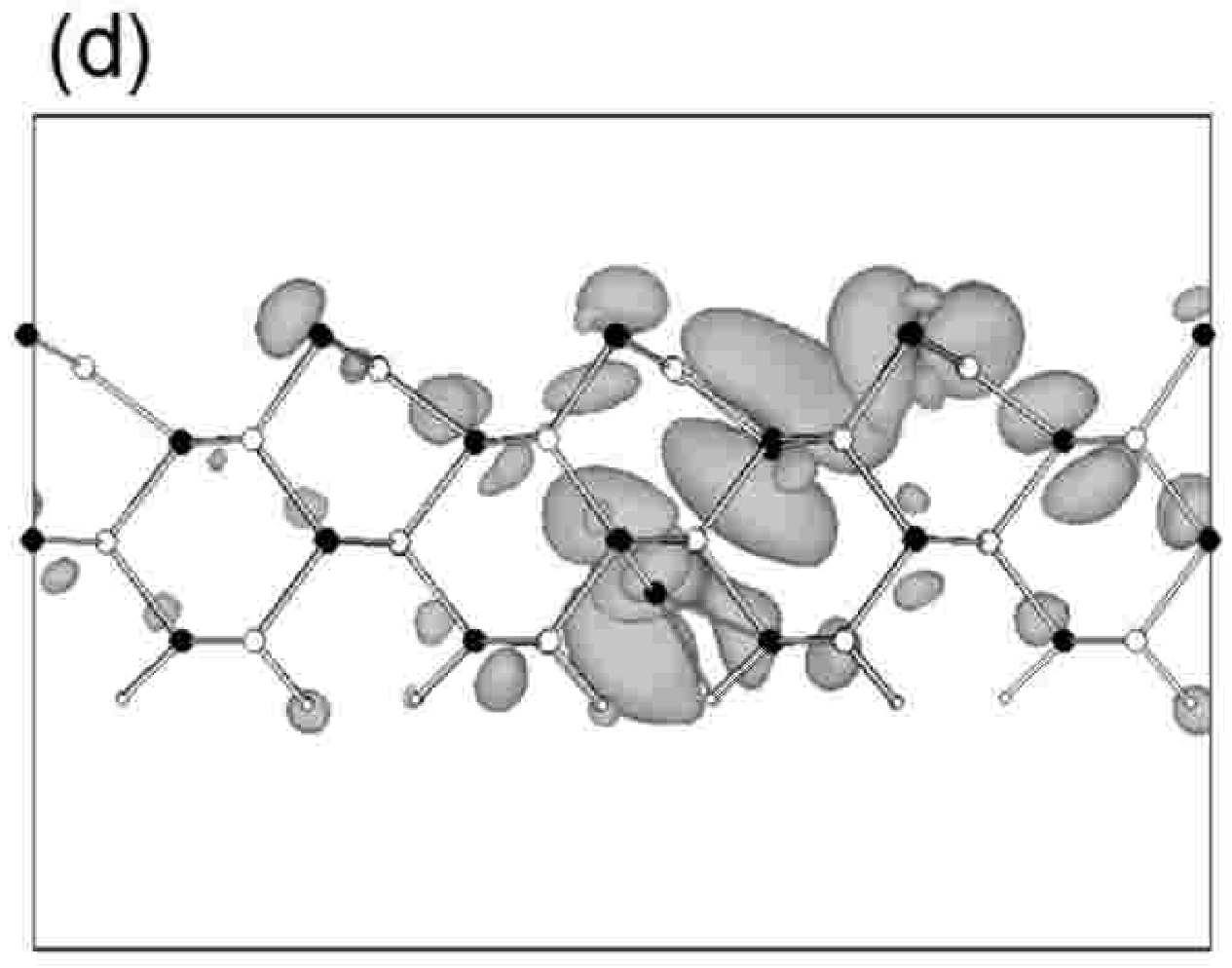}}\\
  \resizebox{60mm}{!}{\includegraphics{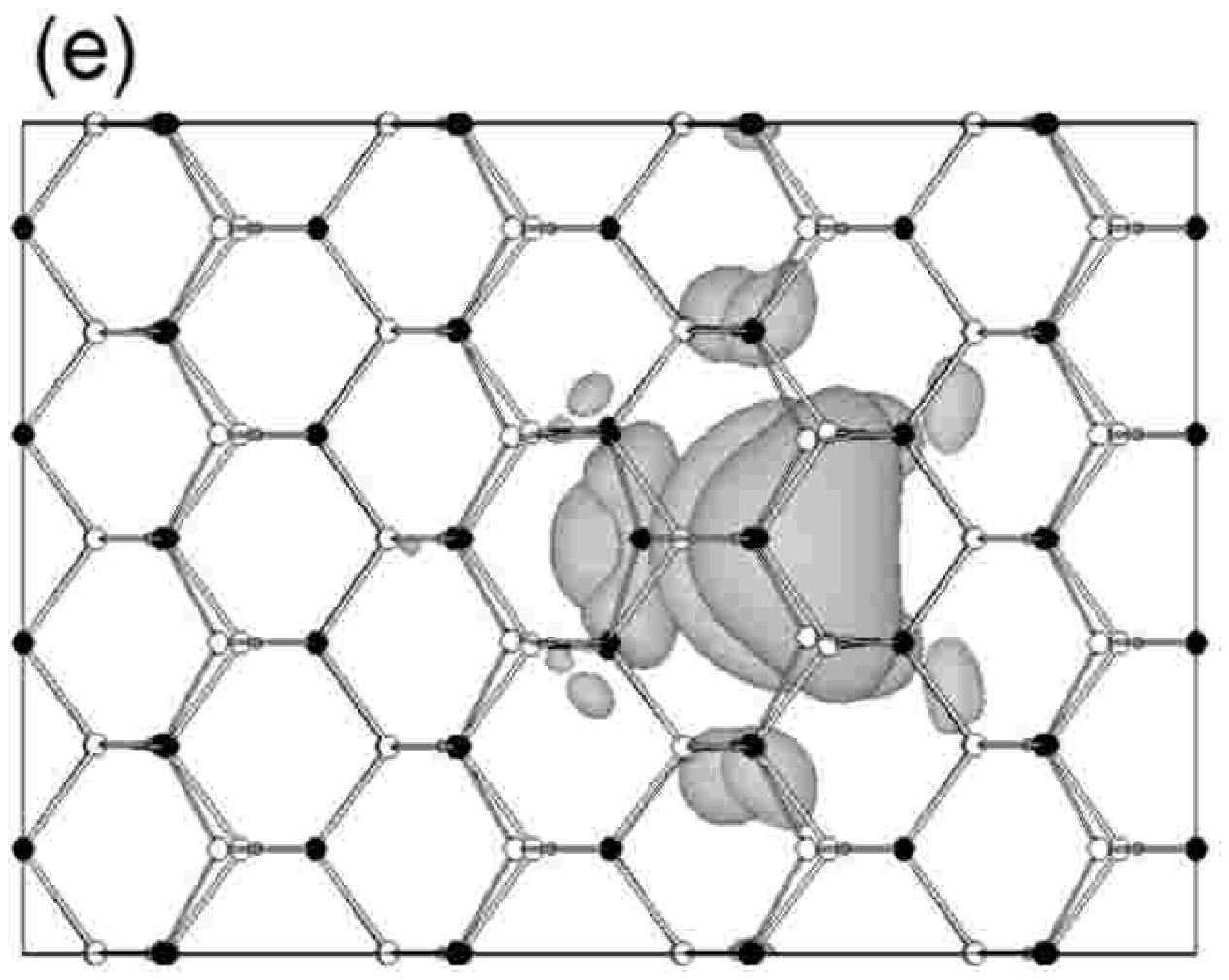}}
  \resizebox{60mm}{!}{\includegraphics{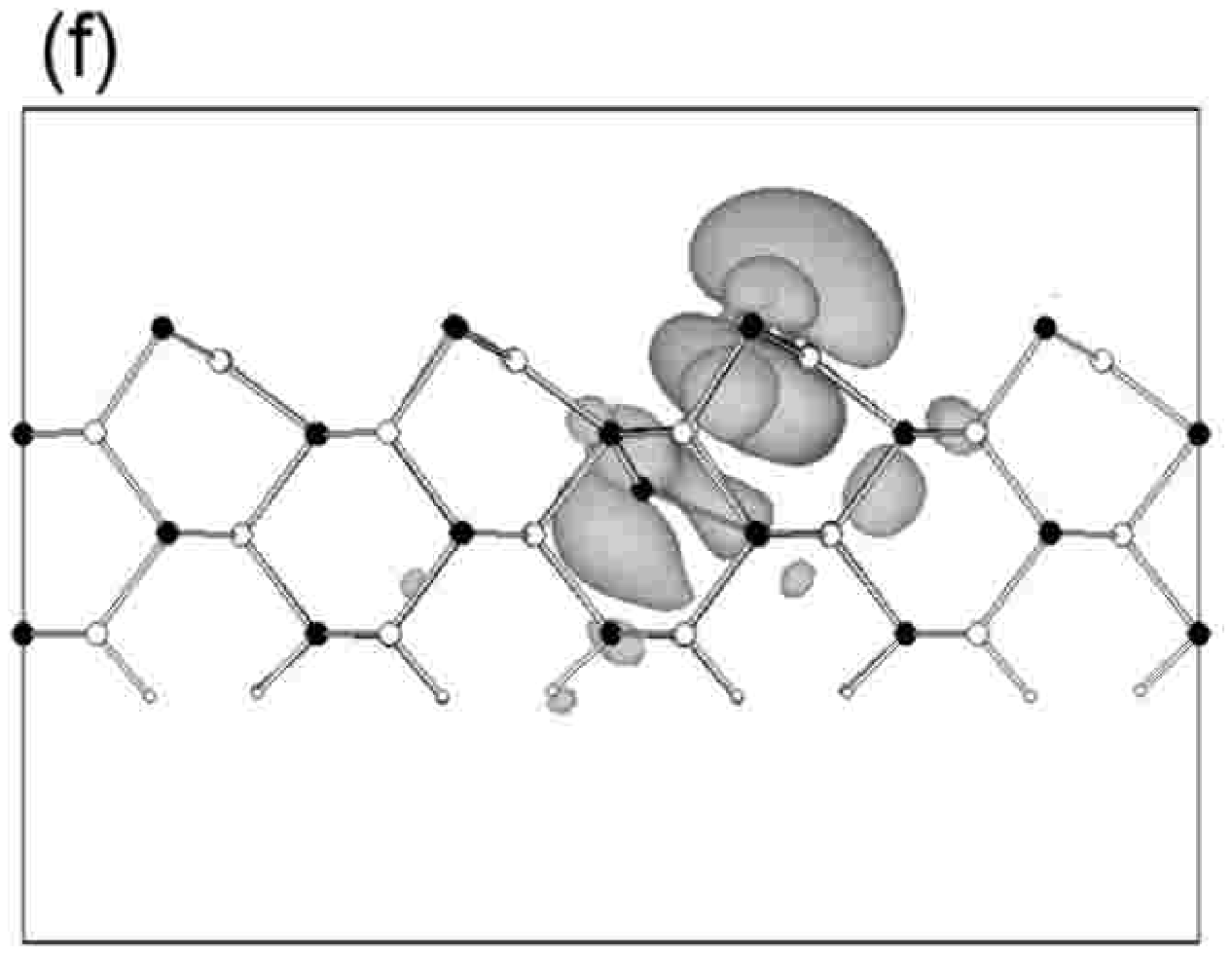}}
  \caption{\label{fig:iso-meta}
The iso-surface of the electron density
when \Asi is located at the metastable position
in the bulk crystal ((a), (b)), in 
the third ((c), (d)) and
the second layer ((e), (f)).
The figures in left and right column are the top 
and side view of the electronic density, respectively, 
as Fig.~\ref{fig:iso-stable}.
In the metastable configuration 
in bulk crystal ((a), (b)), 
there are no nodes 
between the p-orbital of \Asi (colored in red) and 
p-orbitals (colored in gray) of neighboring three As atoms. 
See also the caption of Fig. 4. 
}
\end{figure*}
%---------------------------------------------------------

%----------------------------------------------------------------------%
%                    Metastable configuration                          %
%----------------------------------------------------------------------%
\subsubsection{Metastable configuration}

As mentioned in Sec.~\ref{sec:comp}, 
there are three inequivalent directions
to which the antisite As atom 
could be displaced to form a \Asi-\VGa\ pair.
 Since the STM experiment \cite{hida2001}
shows that the defect contrasts
in the metastable state are all symmetric 
with respect to the [001] axis, 
the ``up'' or ``down'' configurations could be a candidate 
for the metastable state.
As shown later in Sec.~\ref{sec:discussion}, 
only the ``down'' configuration 
gives a better agreement of the calculated STM image 
with that of experiments.
Therefore, only the ``down'' configuratin is discussed
for a possible metastable state of \Asi-\VGa\ pairs
in this section.
When the As antisite is located just on the surface, 
no metastable configuration is found in simulation.

Figures \ref{fig:iso-meta}(c)-(f)  show 
the iso-surfaces 
of the electron density 
of the highest occupied state
 when the \Asi-\VGa\ pair is formed in the layer 3 and the layer 2.
It is notable that the electron density 
increases exclusively along the direction 
opposite to the displacement direction of the \AsGa, 
which is essentially the same as in the bulk lattice. 
The iso-surface localized on the \Asi and the neighboring 
three As atoms in the puckered configuration 
has commonly p-like lobes bonding with each other without nodes 
as in the bulk lattice. 
However, the degree of localization of the highest occupied state 
is slightly dependent on the depth of the \Asi atom. 
When the \Asi atom is located in the layer 2 
(Figs.~\ref{fig:iso-meta}(e) and \ref{fig:iso-meta}(f)), 
the highest occupied state is well localized: 
The electron density is distributed mainly along 
the [11$\overline{1}$] direction.
When the \Asi is located in the layer 3 
(Figs.~\ref{fig:iso-meta}(c) 
and ~\ref{fig:iso-meta}(d)), 
the electron density is less localized having a distribution 
over atoms outside the branches. 
Except for these fine details, the crystal surface has 
little effect on the electronic structure 
of the metastable configuration as in the stable one
as far as the \Asi atom is located below the layer 1.

%%%%%%%%%%%%%%%%%%%%%%%%%%%%%%%%%%%%%%%%%%%%%%%%%%%%%%%%%%%%%%%%%%%%%%%%
%                             Discussion                               %
%%%%%%%%%%%%%%%%%%%%%%%%%%%%%%%%%%%%%%%%%%%%%%%%%%%%%%%%%%%%%%%%%%%%%%%%
\section{\label{sec:discussion}Discussion}

%======================================================================%
%                     Validity of The Embedded Systems                 %
%======================================================================%
\subsection{VALIDITY OF EMBEDDED SYSTEMS}

In constructing the atomic configuration,
two approximations are employed.
First, we relaxed only the atoms in smaller supercell
and fixed the lattice constant.
The expansion of the lattice constant 
by the introduce of \AsGa\
is 1~\%. \cite{staab2001}
It is relatively small in comparison with
the change of the distance 6~\%
between \AsGa\ and its nearest neighbor.
The atomic configuration
around As interstitial is well converged 
in the system of 65 atoms, \cite{staab2001}
and, therefore, we presume 
that it is also the case in \AsGa.
Although we actually tried 
to calculate the electronic structure 
without the relaxation of the atoms,
there remains the features of the electronic structures 
discussed in Figs. 4 and 5.
From these considerations,
we conclude that little error is introduced 
by fixing the lattice constant.

One may consider that the calculated wavefunction
could be sensitive to the number of layers.
We calculated the electronic structure 
of the system with smaller surface area and more layers, 
and  found that the essential characteristics of wavefunctions 
are already converged.
Therefore, the electronic structure is not sensitive 
to the number of layer, or the surface size.

%======================================================================%
%                         Simulated STM images                         %
%======================================================================%
\subsection{SIMULATED STM IMAGES}

%----------------------------------------------------------------------%
%               STM image in the stable configuration                  %
%----------------------------------------------------------------------%
\subsubsection{STM image in the stable configuration}

Figures \ref{fig:stm-stable}(a), \ref{fig:stm-stable}(b),
and \ref{fig:stm-stable}(c) show the simulated 
filled-state STM images of stable 
\AsGa atoms in different depths from the surface.
 The images were calculated by assuming a negative sample bias 
and hence considering 
only the iso-surface of the highest occupied state. 
The image of the \AsGa in the layer 2 
is in good agreement with the previous results.~\cite{capaz1995} 
As mentioned in Sec.~\ref{sec:elec}, even when the \AsGa 
is located near the surface, 
the wavefunction (iso-surface) of the highest occupied state 
does not differ much from that in the bulk lattice 
with the characteristic radial pattern. 
Owing to this fact, 
when the \AsGa is located in the layer 2 
(Fig.~\ref{fig:stm-stable}(a)), 
the STM contrast of the defect core arises 
from the wavefunction concentrated at As atoms 
in the surface layer (the layer 1)
that are neighboring to the \AsGa, 
and the satellite contrasts near the core from the wavefunction 
at the third As neighbors also in the surface layer (the layer 1).
 When the \AsGa is located in the layer 3
(Fig.~\ref{fig:stm-stable}(b)), 
the highest occupied state has a large density 
on the surface just above the \AsGa atom 
and the fifth As neighbors in the surface layer,
giving the satellite features.
 For the \AsGa located in the layer 4 
(Fig.~\ref{fig:stm-stable}(c)), 
the satellite peaks arise from the seventh As neighbors.
 Thus, since the satellite peaks originate in the radial 
pattern of the iso-surface of the highest occupied state, 
they become more remote from the core contrast 
with increasing depth of the \AsGa from the surface.

Feenstra et al. \cite{feenstra1993} argued 
(later referred to as FWP)
that the defect in the STM contrast in their largest size (type A) 
should be assigned to an \AsGa in the layer 1, 
the defect in the second size (type B) to one in the layer 2, 
the third size (type C) to one in the layer 3, 
and the fourth size (type D) to one in the layer 4.
 Their criterion used for judging the depth of the defect 
was the relative position of surface As atom contrasts 
with respect to the defect core contrast.
 Geometrically, if the \AsGa atom is located 
in an odd-numbered layer, 
the mirror symmetric (1$\overline{1}$0) plane 
passing the \AsGa atom must 
cut halfway between As atoms in the surface layer.
 In contrast, if the \AsGa is located in an even-numbered layer, 
the mirror plane must pass an As atom in the surface layer.
 The assignment by FWP of type B defect as an \AsGa in the layer 2 
agrees with the assignment by Capaz et al.~\cite{capaz1995} 
 As for an \AsGa in the layer 1,
Ebert et al. reported a defect image 
that is quite different from type A defects by FWP, 
though such defects were not observed in our experiments probably 
due to the excessively large magnitude 
of the bias voltage.~\cite{ebert2001}
Figures \ref{fig:stm-stable}(d) and 
\ref{fig:stm-stable}(e) 
show experimental STM images typically observed 
at a sample bias of $-2.3$~V.
 Our experimental images shown 
in Figs.~\ref{fig:stm-stable}(d) 
and \ref{fig:stm-stable}(e) seem 
identical to respectively  the type B and C defects
in experiments by FWP.~\cite{feenstra1993}

Another criterion for depth assignment 
is the distance between the two satellites 
from the core contrast as stated above.
 The order of the experimental images 
in Fig.~\ref{fig:stm-stable} 
are tentatively arranged according to this criterion.
 Although the quantitative agreement is not perfect, 
we could assign the image shown 
in Fig.~\ref{fig:stm-stable}(e) to an \AsGa in the layer 3, 
and image Fig.~\ref{fig:stm-stable}(d) to an \AsGa 
in the second layer.
 The assignment of the type B defect (Fig.~\ref{fig:stm-stable}(d)),
the same as proposed by Capaz et al.,~\cite{capaz1995}  
seems reasonable 
because the agreement of the defect contrast 
in the metastable state is better 
than otherwise as shown in what follows.

%----------------------------------------------------------------------%
%               STM image in the metastable configuration              %
%----------------------------------------------------------------------%
\subsubsection{STM image in the metastable configuration}

\begin{figure}
  \resizebox{35mm}{!}{\includegraphics{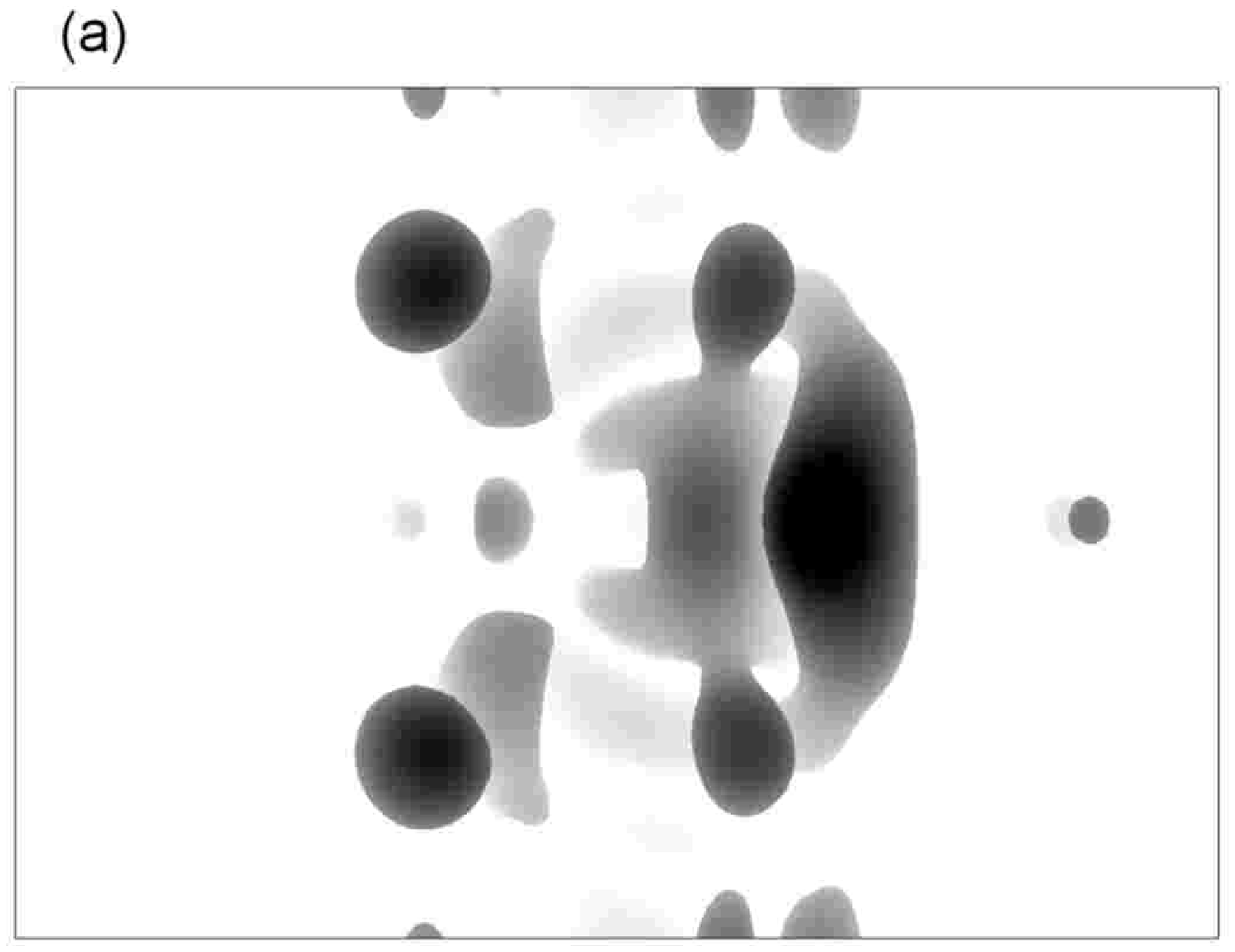}}
  \resizebox{35mm}{!}{\includegraphics{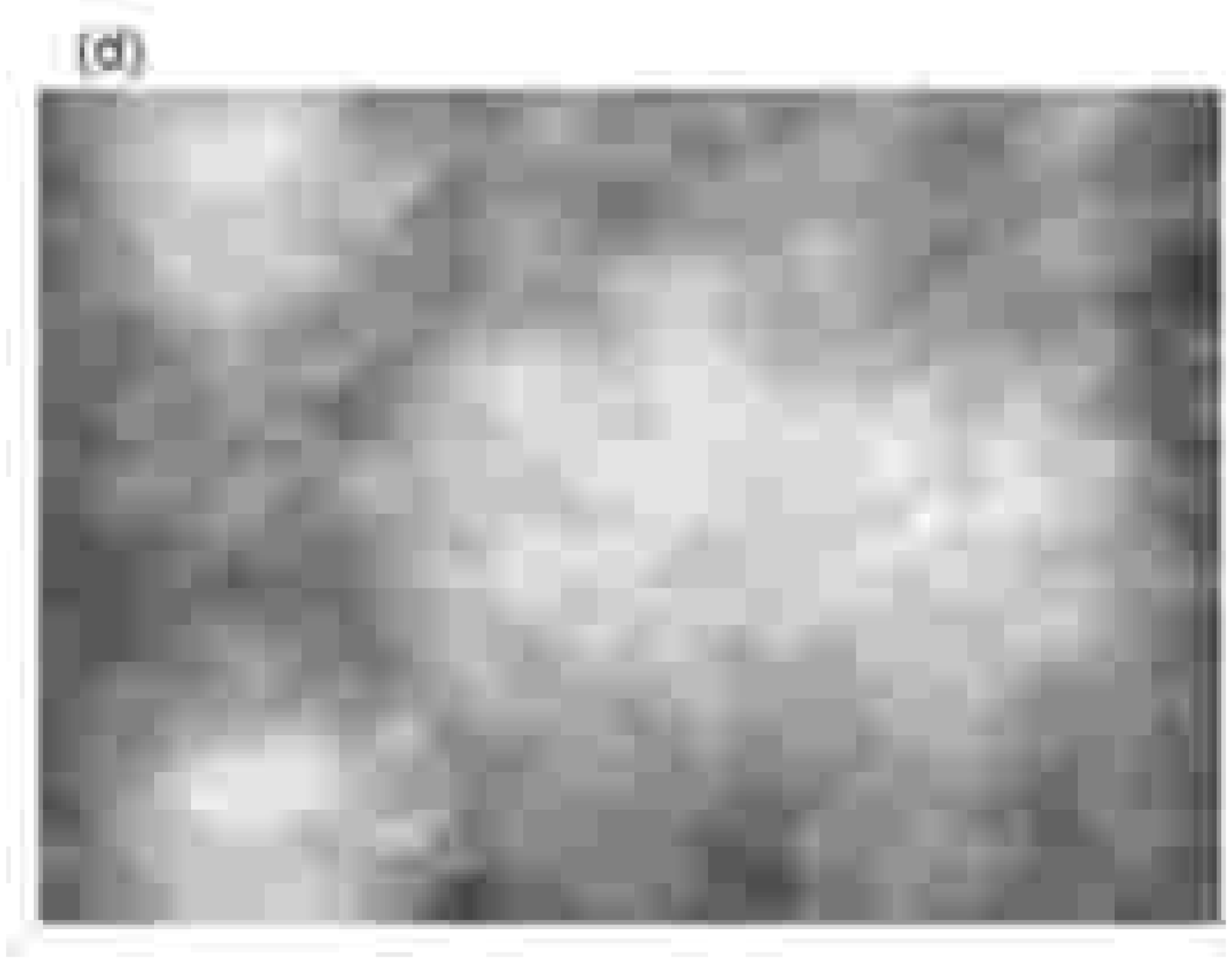}}\\
  \resizebox{35mm}{!}{\includegraphics{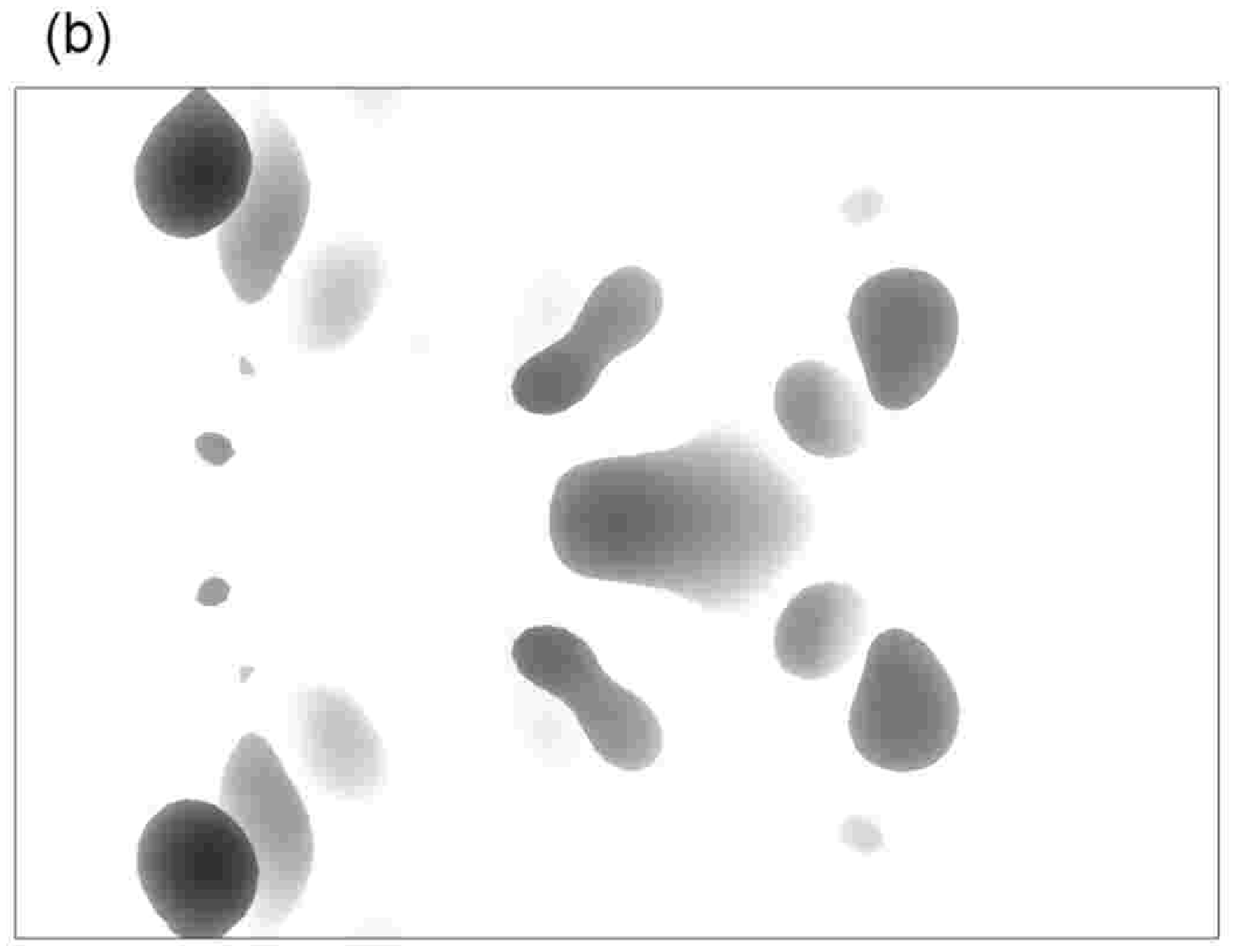}}
  \resizebox{35mm}{!}{\includegraphics{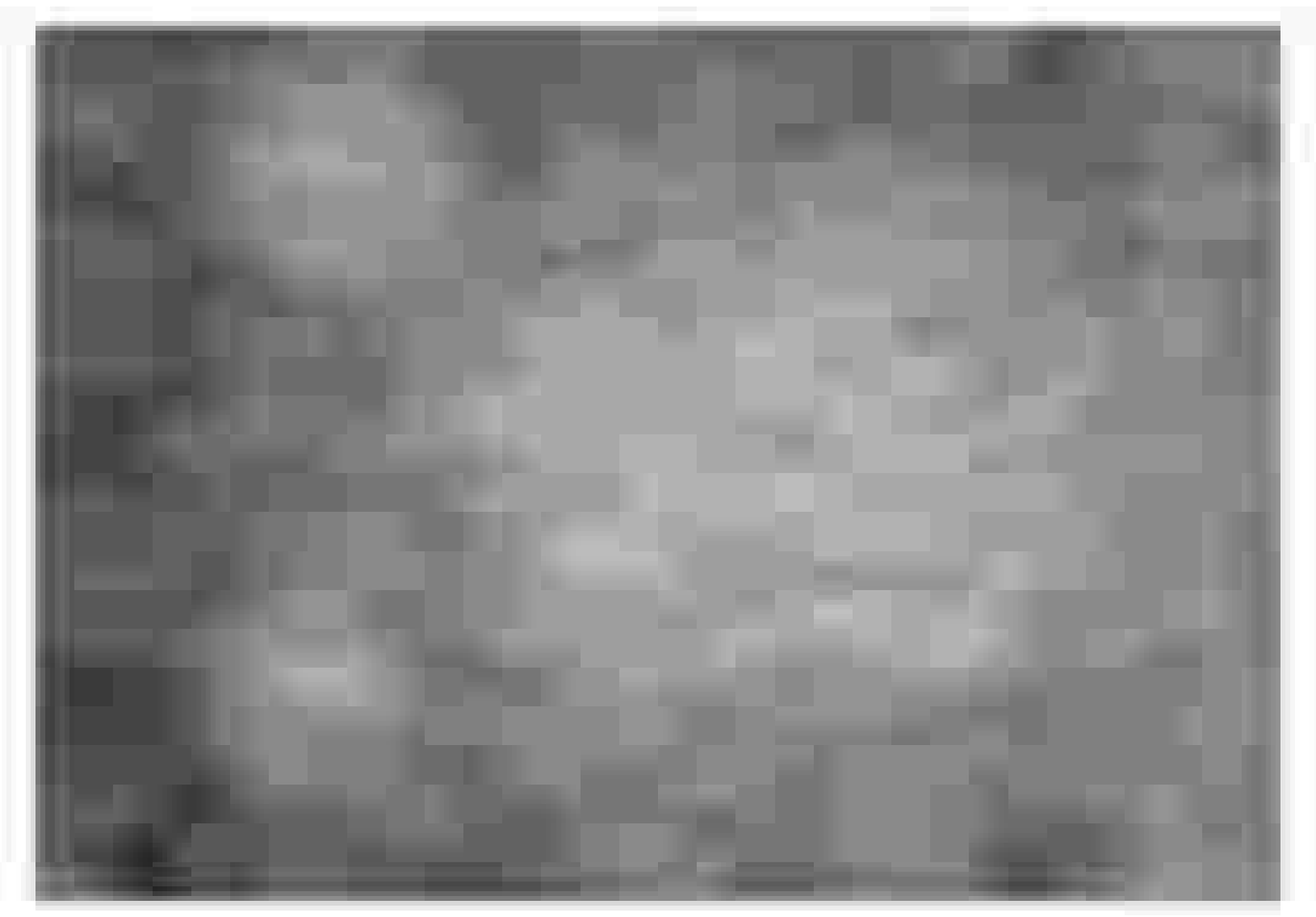}}\\
  \resizebox{35mm}{!}{\includegraphics{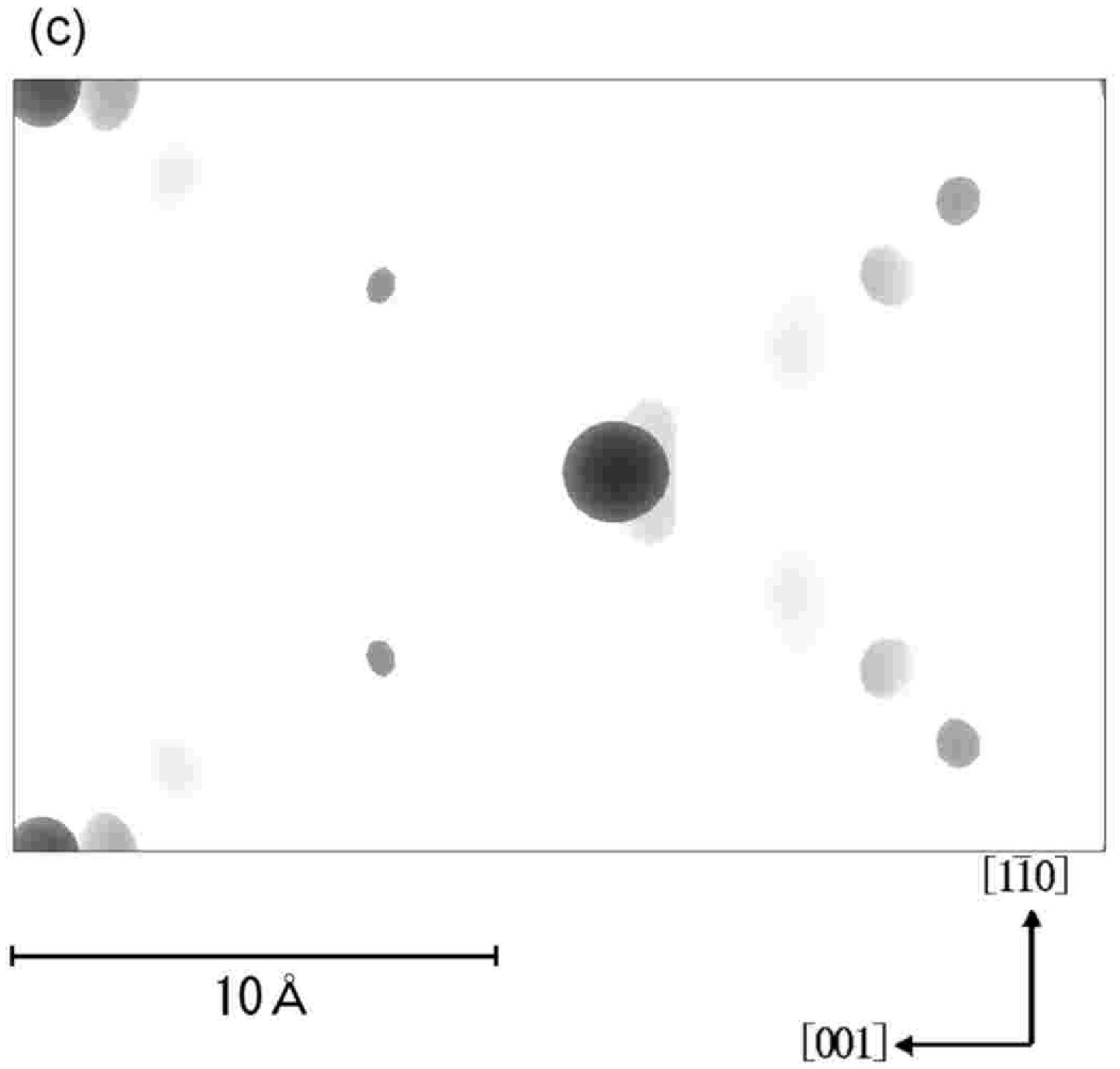}}
  \resizebox{35mm}{!}{\hspace{35mm}}
  \caption{\label{fig:stm-stable}
 The iso-surface of the electron density $8.32\times 10^{-4}$e/\AA${}^3$
of the highest occupied band when \AsGa locates
near the crystal surface.
 The \AsGa locates in 
(a) the layer 2,
(b) the layer 3, and
(c) the layer 4.
 The color becomes darker with the iso-surface approaching the vacuum.
 It can be seen that the distance between two satellite peaks becomes larger 
as \AsGa locates at deeper positions.
Figures (d) and (e) are experimental STM images.}
\end{figure}

\begin{figure}
  \resizebox{35mm}{!}{\includegraphics{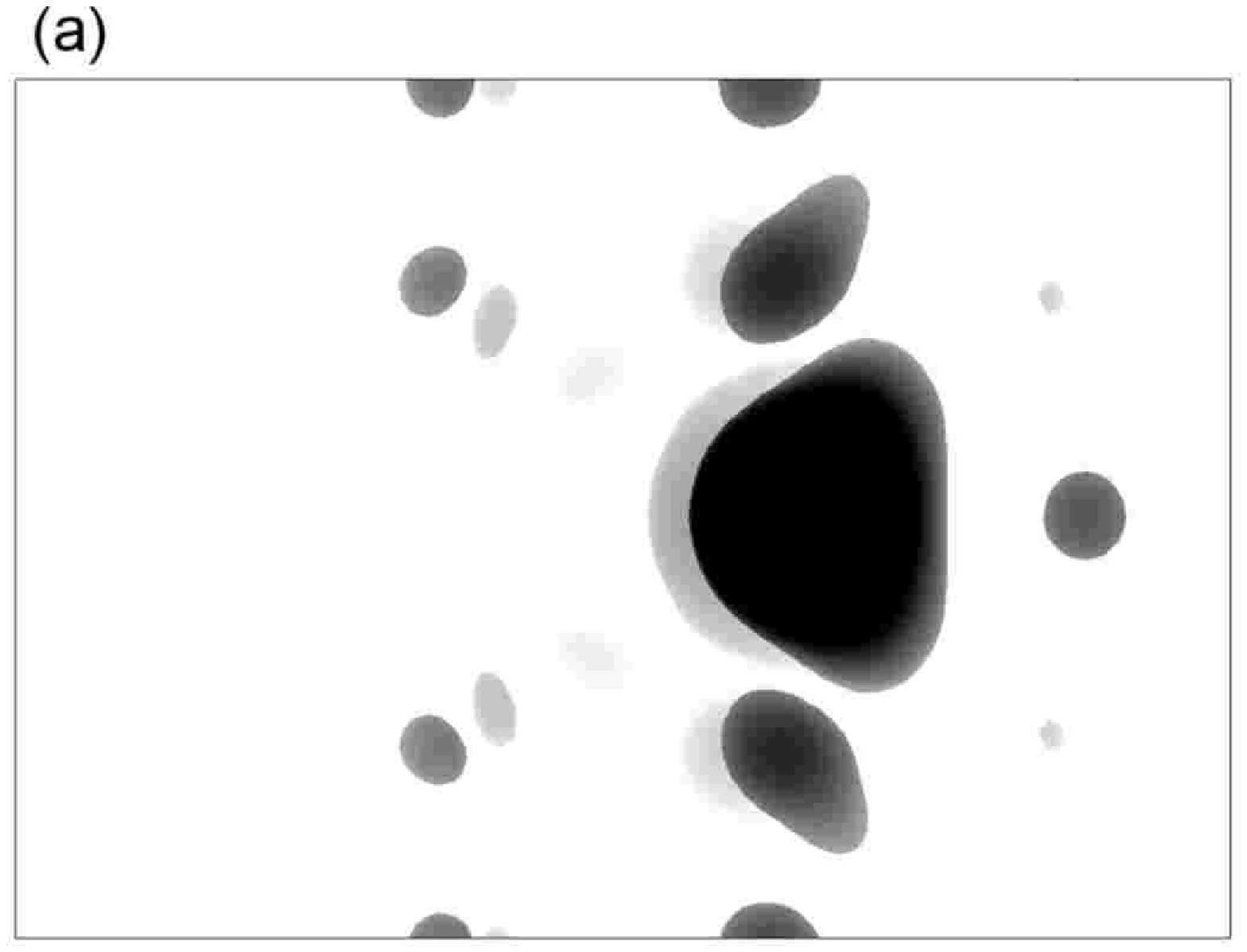}}
  \resizebox{35mm}{!}{\hspace{35mm}}\\
  \resizebox{35mm}{!}{\includegraphics{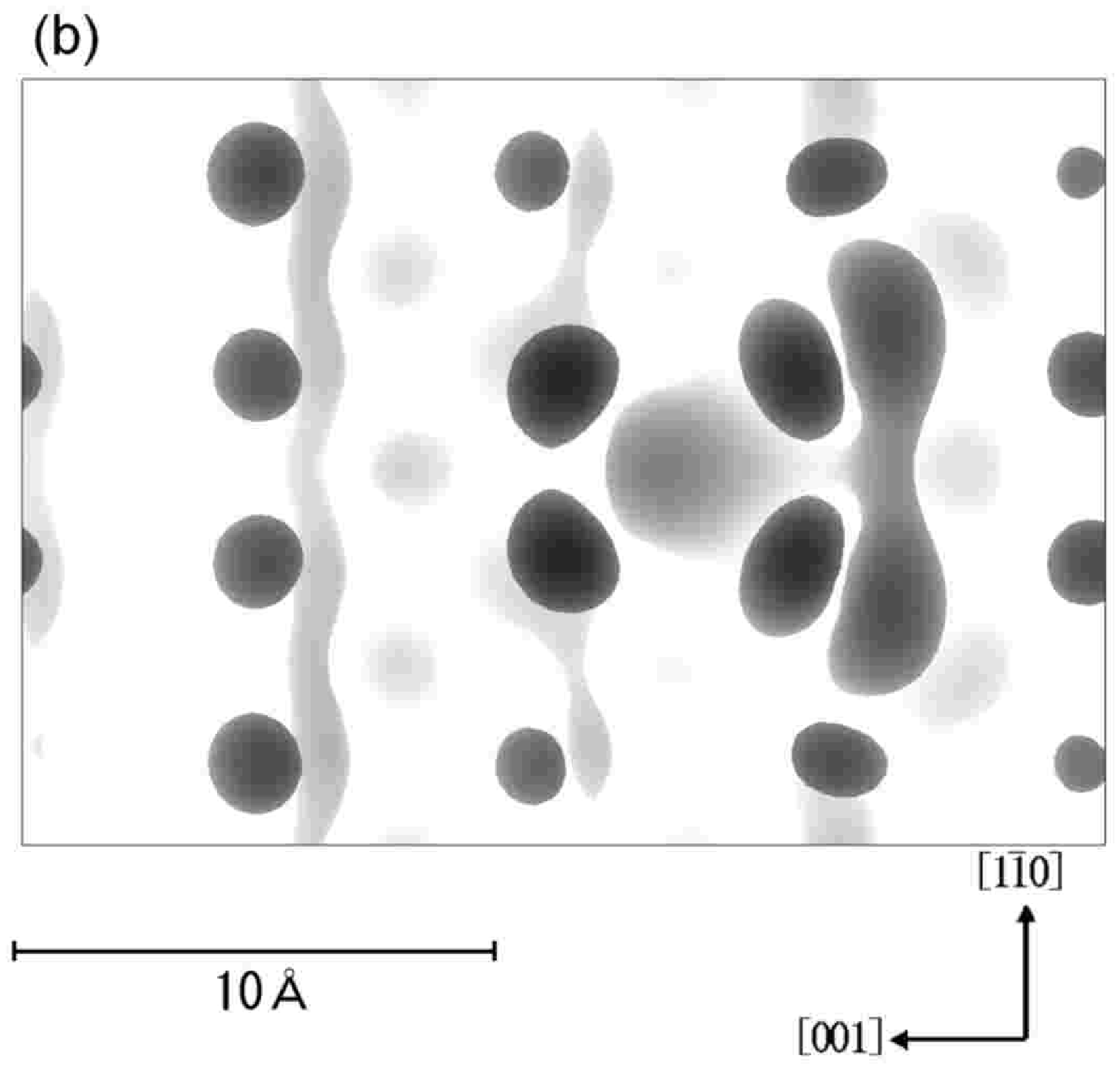}}
  \resizebox{35mm}{!}{\includegraphics{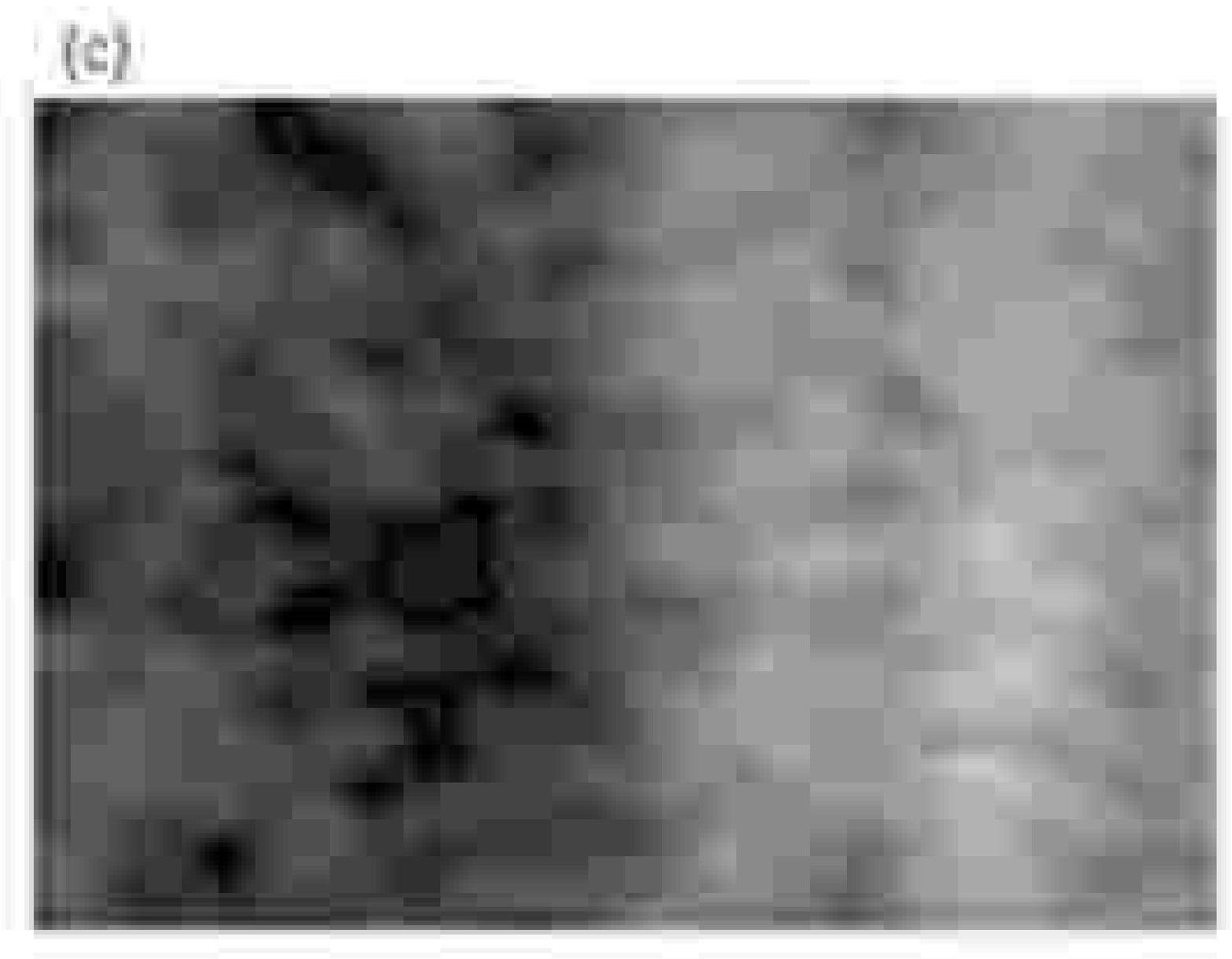}}\\
  \caption{\label{fig:stm-metastable}
The iso-surface of the electron density of the highest occupied band
when \Asi locates at metastable positions in 
(a) the layer 2 and
(b) the layer 3.
The satellite peaks in the stable configuration vanish 
(Compare  with Fig.~\ref{fig:stm-stable}).
Figure (c) is an experimentally observed STM image,
of defects as Fig.~\ref{fig:stm-stable}(e) 
in the metastable state.}
\end{figure}

 Figure~\ref{fig:stm-metastable}(c) 
shows the experimental STM image 
of the defect in the metastable state 
that underwent a change from the contrast shown 
in Fig.~\ref{fig:stm-stable}(e).
 The image in the metastable state is characterized 
by the diminish of the satellite peaks 
and the concomitant appearance of a new contrast 
elongated along the surface As atom along [110] row 
on the side opposite to the diminished satellites.
As shown in Sec.~\ref{sec:elec}, 
the branches of the radial wavefunction of the highest occupied state 
disappear upon transformation to the metastable configuration, 
except for only a branch extending in the [111] direction 
opposite to the displacement of the \AsGa atoms.
 This means that if the \AsGa atom near the surface 
was displaced to a ``side'' position, 
the STM image would lose the mirror symmetry 
about the (110) plane.
 If the \AsGa atom was displaced to the ``up'' position, 
the peak position move in the opposite direction 
to the one experimentally observed.
 Both of these contradict the experiments.~\cite{hida2001}

 This conclusion is different from the calculations 
by Zhang,~\cite{zhang1999}
where the ``side'' configuration in the second layer
is the most stable energetically 
among three possible configurations of the metastable state. 
Our calculation shows the similar energy barrier 
in the path of \AsGa\ 
from the metastable to the stable positions in these
configurations.
The energy barrier heights are 0.5 eV in  
the ``side'' configuration in the second layer, 
0.3 eV in the ``down'' configuration in the second layer 
(Fig.~\ref{fig:stm-metastable}(a)), 
and 0.3 eV in the ``down'' configuration in the third layer 
(Fig.~\ref{fig:stm-metastable}(b)).
 Therefore, it is still a open question
why only one metastable defect image 
has been observed in STM experiment 
that is not energetically favorable.

 Figures~\ref{fig:stm-metastable}(a) 
and \ref{fig:stm-metastable}(b) show 
the simulated STM images of \Asi-\VGa\ pairs located 
in the layer 2 and in the layer 3, 
respectively, with the \Asi\ in the ``down'' configuration.
 The agreement between simulation and experiment is better 
if the \AsGa is assumed to be located in the layer 3 
as tentatively assigned above. 

%======================================================================%
%                        Origin of Metastability                       %
%======================================================================%
\subsection{Origin of metastability}

The physical origin of the metastability of
\Asi-\VGa\ pairs can be understood 
if one examines the change in the bonding character 
upon the stable to metastable transformation.
 As shown in Sec.~\ref{sec:elec},
the \AsGa\ configuration, though lower in total energy, 
bears an inherent instability due to the anti-bonding nature 
of the chemical bonds with the surrounding atoms.
 The global expansion of the lattice by the presence of 
\AsGa\ defects is, thus, owing to this anti-bonding nature 
of the electronic structure associated 
with the defects.~\cite{staab2001} 
  On transformation to the metastable state, 
the electronic energy of the s-orbital associated 
with the \AsGa\ atom that is anti-bonded 
with the nearest As atoms is lifted and becomes emptied.
 In its place, 
a gap state associated with an atomic orbital 
of p-character originating in the displaced As atom 
is drawn from the conduction band, 
and it becomes occupied to form bonds 
with three neighboring As atoms.
 Thus, some of the increase of the total energy 
on transformation is canceled by the newly formed chemical bonds 
which contributes to the affinity 
of the interstitial As atom with the surrounding.

Since the energy gain by such bond reconstruction 
should become larger with decreasing distance 
between the atoms, 
the change of the dominant orbital of the \AsGa 
from s to p-orbital, 
and the disappearance of the anti-bonding node 
between \Asi and the surrounding As atoms 
approached by \AsGa\ are considered to be 
the reasons why the metastable state exists.

%======================================================================%
%                     As antisite in (110) surface                     %
%======================================================================%
\subsection{As antisite in (110) surface}

As shown in Figs.~\ref{fig:iso-stable} and ~\ref{fig:iso-meta}, 
the presence of surface has no significant effects 
on the electronic structures associated with \AsGa\ defects 
as far as they are located deeper than the layer 2.
It is only when the \AsGa is located in the surface layer 
(the layer 1) 
that no level is formed in the band gap.
 Due to the absence of the gap state
or the spatial spread of the occupied states, 
\AsGa\ defects in the layer 1 give rise 
to no localized STM contrast at negative sample biases, 
in agreement with the results obtained by Ebert et al. \cite{ebert2001}

 The surface buckling \cite{alves1991} 
which is reproduced in the present calculation as well is induced 
by the electron transfer in the Ga dangling bond 
which is higher in energy to the As dangling bond.
 This causes a change of the orbital character 
from $\mathrm{sp^3}$ in the perfect lattice to $\mathrm{sp^2}$-like 
on the surface Ga atom and to $\mathrm{p^3}$-like 
on the surface As atom,
thereby resulting in the retraction 
of the Ga atom and the protrusion of the As atom from the surface.
 Concomitantly, the As dangling bond state smears into the valence band, 
and the Ga dangling bond state into the conduction bands.
It is expected that when \AsGa\ defect is located 
in the surface (layer 1),
\AsGa reverts to the unbuckling position
as noticed in a figure of the paper 
by Ebert et al. \cite{ebert2001}

\begin{figure}
  \resizebox{40mm}{!}{\includegraphics{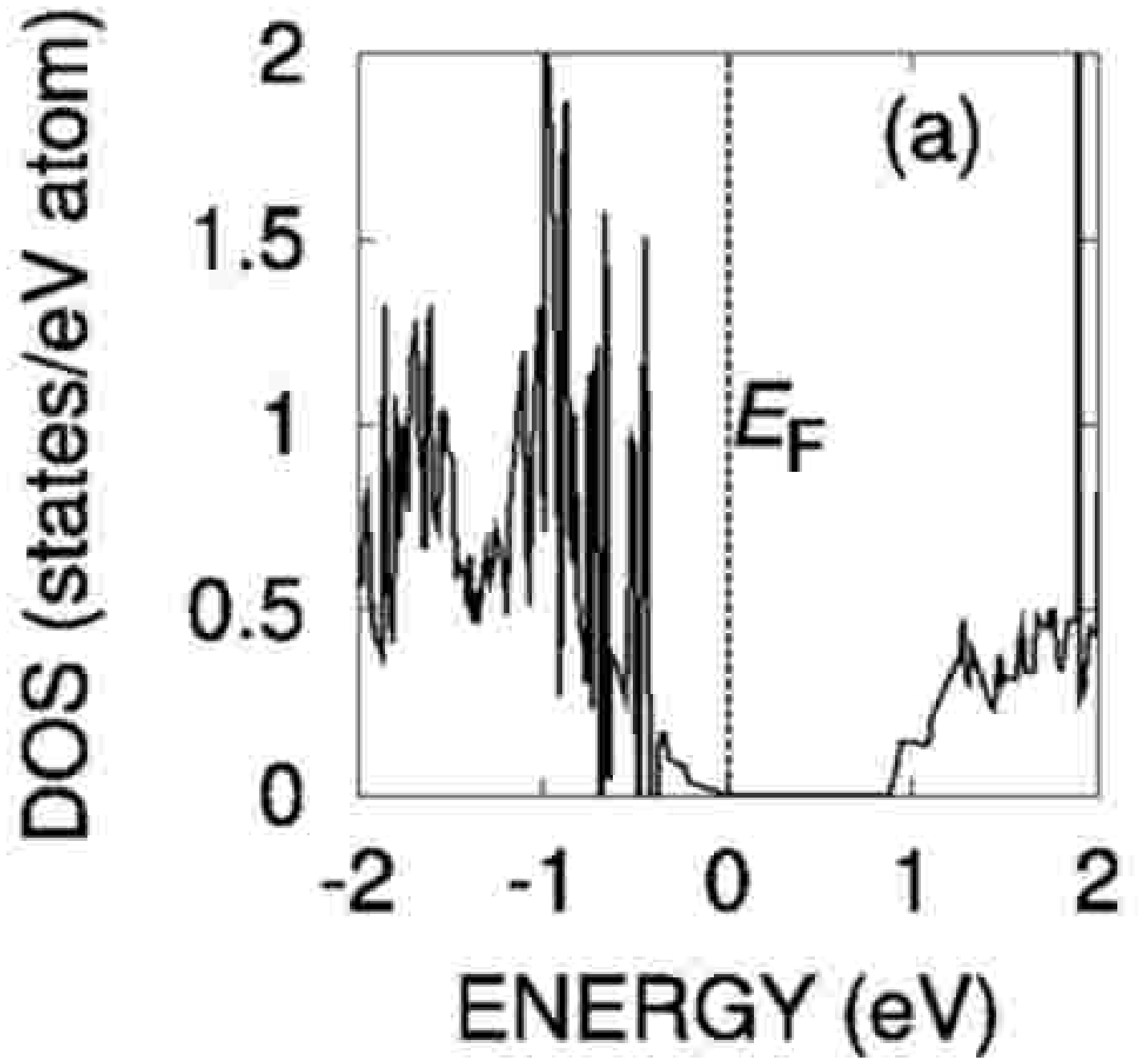}}
  \resizebox{40mm}{!}{\includegraphics{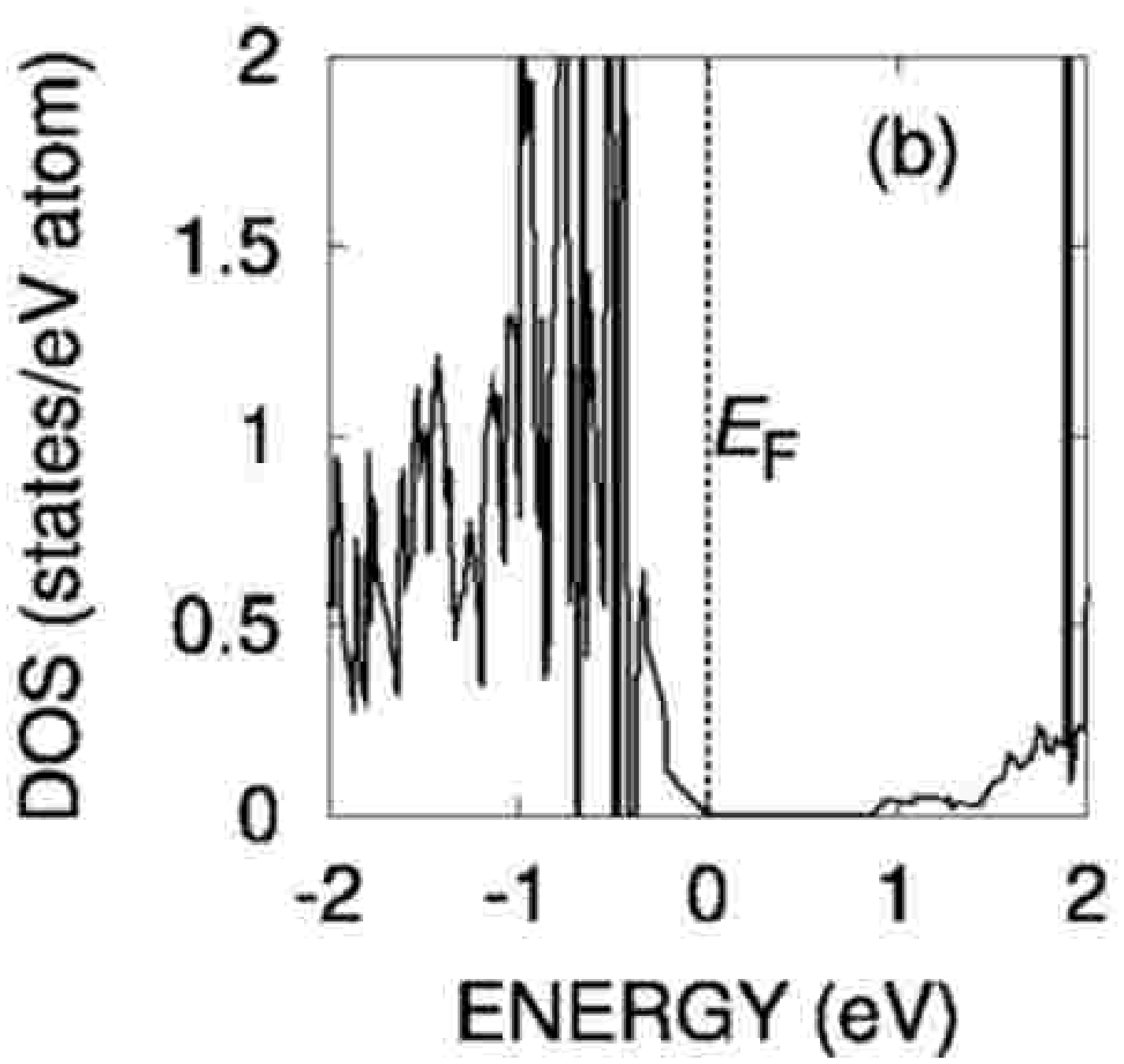}}
  \caption{\label{fig:stm-l0-dos}
  The local density of states (\lDOS) of
(a) the As atom in the surface layer next to \AsGa in the surface layer, 
and 
(b) an As atom in the surface layer further away from \AsGa.
The weight of \lDOS\ just below the Fermi energy 
(between $E_{\rm F}-0.4$~eV and $E_{\rm F}$),  
of (a) is smaller than that of (b).
} 
\end{figure}

To test this idea,
we calculated a model system of 60 atoms 
with the surface size $2 \times 3/\sqrt{2}$ and 
with 5 layers including the hydrogen layer 
by the LMTO method.
 This model had a clean (110) surface on one end, 
and the other surface was terminated by hydrogens.
 Then, \AsGa\ was positioned at the same height of As atoms 
in the surface layer.
 Figures \ref{fig:stm-l0-dos} shows the \lDOS\
at the As atom in the surface layer next 
to the \AsGa (a) and As atoms distant from the \AsGa (b).
   The height of the \lDOS\ just below the Fermi energy 
(between $E_{\rm F}-0.4$~eV and  $E_{\rm F}$) 
is quite different;
As atoms in the surface layer away from \AsGa contain electrons 
more than an As atom next to \AsGa.
 This means that the electrons on the As atoms next to \AsGa 
in the surface layer transfer into \AsGa.

\begin{figure}
  \resizebox{40mm}{!}{\includegraphics{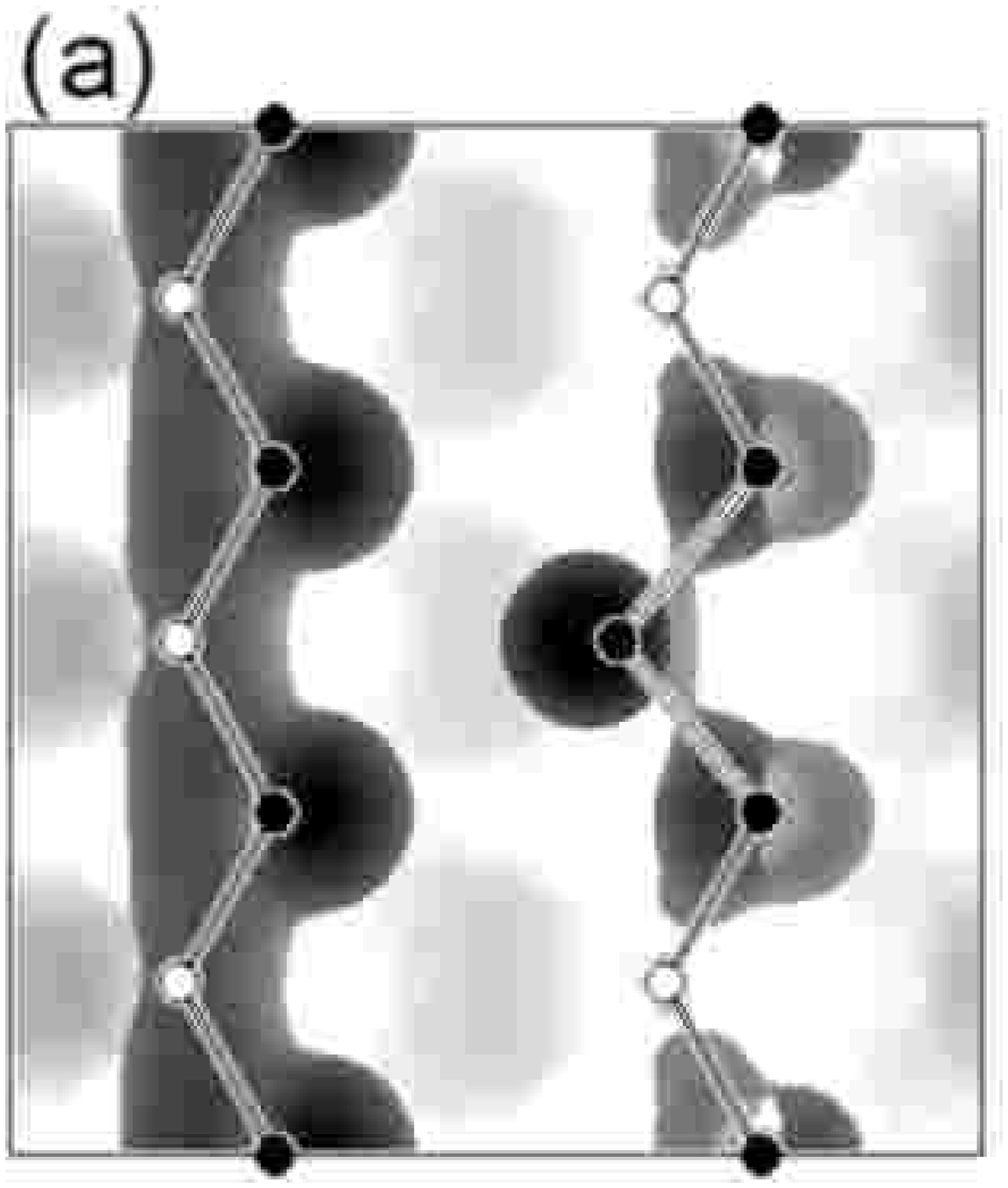}}
  \resizebox{40mm}{!}{\includegraphics{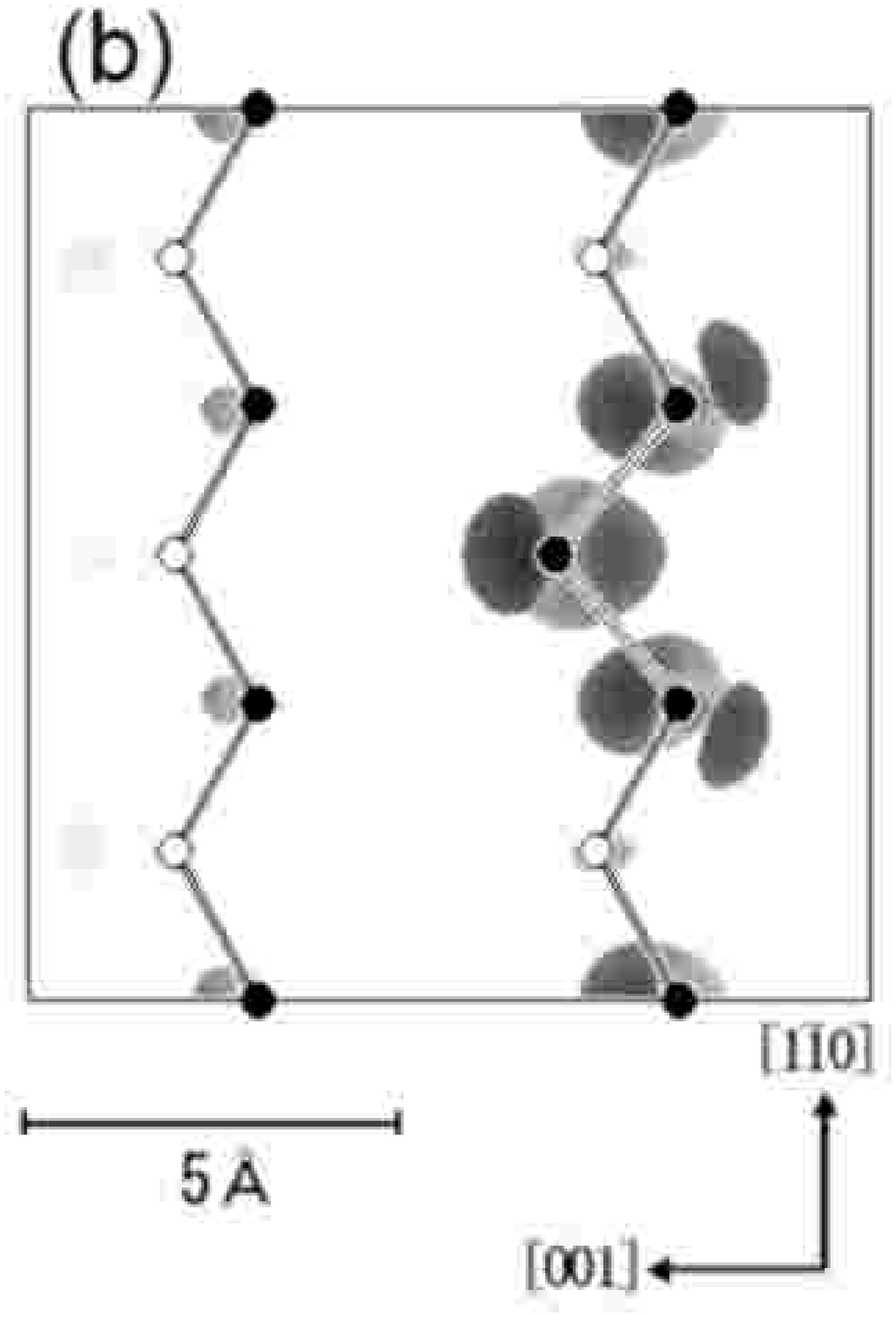}}
  \caption{\label{fig:stm-l0}
 The iso-surface of the electron density on the crystal surface
of the highest occupied band calculated by the LMTO method
when an \AsGa is in the surface layer. 
(a) Between the top of valence band and 0.2~eV below, and
(b) between the bottom of conduction band and 0.2~eV above it.
 The \AsGa is located at the center of the right zigzag chain.
} 
\end{figure}

 Atoms in the surface layer align 
in one-dimensional zigzag chains, 
but each chain does not connect with adjacent chains.
Figure~\ref{fig:stm-l0}(a) shows
the iso-surface of the highest occupied state 
which is extended over the whole system 
with the major component along the [1$\overline{1}$0] chain 
of surface As atoms 
next to the [1$\overline{1}$0] chain containing the \AsGa.
 The large electron density on this chain originates 
mainly in the dangling bond orbitals of the As atoms.
 The much smaller but some density on the \AsGa 
is due to the dangling bond p-orbital of the \AsGa atom.
 The electron density on the other As atoms 
in the [1$\overline{1}$0] surface chain 
containing the \AsGa does not originate
in the dangling bond 
but in the As p-orbitals parallel 
to the dangling p-orbital of the \AsGa.
 In contrast to the highest occupied states, 
the lowest unoccupied state (Fig.~\ref{fig:stm-l0}(b)) 
is localized around the \AsGa.
 The iso-surface consists mainly of the dangling bonds 
of the \AsGa and the As atoms in the chain containing the \AsGa.
 In other words, 
the dangling bond orbitals of the As atoms are occupied 
when \AsGa is absent in the zigzag chain, 
and are unoccupied when \AsGa is present in the chain.
 This means 
that electrons transferring to \AsGa are 
those occupying the dangling bond orbitals of As atoms 
next to \AsGa in the surface
in case of the perfect lattice.
 When \AsGa located in the surface layer, 
electron transfer does not occur  
because the energy of dangling bond orbital of 
As is almost the same 
as that of \AsGa 
and the buckling does not stabilize 
the state of the electron transfer.
Thus, the surface buckling on the site of \AsGa in the surface layer 
does not take place.
%

%___________________________________________________

%___________________________________________________

%%%%%%%%%%%%%%%%%%%%%%%%%%%%%%%%%%%%%%%%%%%%%%%%%%%%%%%%%%%%%%%%%%%%%%%
%                            Conclusion                               %
%%%%%%%%%%%%%%%%%%%%%%%%%%%%%%%%%%%%%%%%%%%%%%%%%%%%%%%%%%%%%%%%%%%%%%%
\section{\label{sec:conclusion} Conclusion}

 We presented the electronic structures of the gap state
associated with a single \AsGa in GaAs.
 The wavefunction in the stable configuration mainly consists
of the s-orbital of \AsGa
and the p-orbitals of the surrounding As atoms.
 These p-orbitals spread
around \AsGa with a radial pattern 
of aligned of As p-orbitals.
 On the other hand, in the metastable configuration,
the major component of the wavefunction is
the p-orbital of the \Asi and the As atoms around it.
 In comparison with the stable configuration,
some p-orbitals of the first neighbor As atoms 
changes their heading directions
breaking the radial pattern of the alignment of the As p-orbitals.
 As a result, 
a small part of the radial pattern 
in the stable configuration remains. 
 These transitions of the orbital and the bonding character
can be regarded as the origin of the metastability.   
 These essential features of electronic structures 
do not change
even when the \AsGa is located near the surface.
However, when the \AsGa is located just on the surface,
the surface unbuckles leaving
no localized state around the top of the valence band.
 We can attribute the defect image in the STM experiments
to these wavefunction,
and the disappearance of the satellite peaks of the images to
the change in the wavefunction accompanied
by the change to the metastable state.
 Although they does not thoroughly support  experimental results,
the results of the present calculations provide simple and clear 
explanation for the origins of the metastability and the characteristic 
defect images in the STM experiments.  

%----------------------------------------------------------------
\section*{ACKNOWLEDGEMENTS}
This work was  supported partially by a Grant-in-Aid 
for Priority Area on 
``Manipulation of Atoms and Molecules by Electronic Excitation'' 
from the  Ministry of Education, Culture, Sports, Science 
and Technology (MEXT) of Japan.

\bibliography{el2}

\begin{thebibliography}{25}
\bibitem[{foo()}]{footnote1}
\bibinfo{note}{Although a calculation was done by Zhang \cite{zhang1999} for
  the \Asi -\VGa\ pair as the defect in the metastable state, the bias polarity
  was opposite to that used in the experiments \cite{hida2001}}.
\end{thebibliography}

\end{document}